\documentclass[onecolumn, extra, mreferee]{gji}

\usepackage{graphicx}
\usepackage{amsmath}
\usepackage{amssymb}
\usepackage{lineno}
\usepackage{multirow}
\usepackage[section]{placeins}

\title[EM with conductive, permeable wells]{
  Impacts of magnetic permeability on electromagnetic data collected in settings with steel-cased wells
}

\author[L. J. Heagy and D. W. Oldenburg]{
  Lindsey J. Heagy$^1$ and Douglas W. Oldenburg$^1$\\
  $^1$ Geophysical Inversion Facility, University of British Columbia, Vancouver, BC, Canada, V6T 1Z4
  \\ \quad email: lheagy@eoas.ubc.ca
}

\begin{document}

\label{firstpage}

\maketitle

\begin{summary}

Electromagnetic methods are increasingly being applied in settings with steel infrastructure. These include applications such as monitoring of CO$_2$ sequestration or even assessing the integrity of a wellbore. In this paper, we examine the impacts of the magnetic permeability of a steel-cased well on electromagnetic responses in grounded source experiments. We consider a vertical wellbore and simulate time and frequency domain data on 3D cylindrical meshes. Permeability slows the decay of surface electric fields in the time domain and contributes to a phase shift in the frequency domain. We develop our understanding of how permeability alters currents within, and external to, the casing by focussing first on the time domain response and translating insights to the frequency domain. Following others, we rewrite Maxwell's equations to separate the response into terms that describe the magnetization and induction effects. Magnetic permeability impacts the responses in two ways: (1) it enhances the inductive component of the response in the casing, and (2) it creates a magnetization current on the outer wall of the casing. The interaction of these two effects results in a poloidal current system within the casing. It also generates anomalous currents external to the casing that can alter the geometry and magnitude of currents in the surrounding geologic formation. This has the potential to be advantageous for enhancing responses in monitoring applications.

\end{summary}

\begin{keywords}
Electromagnetic theory, Controlled source electromagnetics (CSEM), Downhole methods, Numerical modelling, Electrical properties, Magnetic properties
\end{keywords}


\section{Introduction}

Steel casing and pipelines are an ever-present part of today's infrastructure. They are used for extracting hydrocarbons, generating geothermal energy, or disposing waste-water. One important and growing application is the use of new and existing wells for geologic sequestration of CO$_2$. In many of these applications, electromagnetic (EM) data have the potential to provide useful insights. In applications where wells are used to inject or extract fluids to depths there are two aspects of relevance. The most common application is to monitor the distribution of fluids with time. Applications in hydrocarbons include reservoir imaging and monitoring of enhanced oil recovery or hydraulic fracturing \citep{Rocroi1985, Pardo2008a, tang_three-dimensional_2015, hoversten_hydro-frac_2015, Tietze2015}. Recently, there has been growing interest in the use of electromagnetics for monitoring CO$_2$ as it is injected in order to monitor for leaks or improve operations \citep{carrigan_electrical_2013, park_25d_2017, puzyrev_three-dimensional_2017, um_joint_2020}. The second major area of application is to evaluate the integrity of a well or pipeline \cite{wilt_casing_2020, beskardes_effects_2021}. A well that is corroded can be a conduit for fluids to migrate from deep reservoirs and contaminate shallow aquifers. A recent special issue of The Leading Edge provides an overview of a range of applications where EM is applied in settings with steel-cased wells \citep{weiss_introduction_2022}.

The presence of steel infrastructure can be a complicating factor for the use of EM in these settings. In a cross-well or surface to borehole survey where magnetic field sensors are deployed in a borehole, steel casing attenuates signals \citep{augustin_theoretical_1989, Wu1994, Wilt1996, cuevas_analytical_2014}. Additionally, steel infrastructure has an EM response that contributes ``noise'' that must be accounted for in numerical simulations or inversions. Although steel casings are a complicating factor for numerical modelling and inversions, multiple authors have shown that they can act as ``extended-electrodes'' that can help excite targets at depth and enhance signals that may not be observable had there been no steel-casing present \citep{schenkel_electrical_1994, hoversten_hydro-frac_2015, yang_3d_2016, puzyrev_three-dimensional_2017}.

Solutions are established for simulating DC resistivity experiments in settings with steel infrastructure \cite{schenkel_electrical_1994, yang_3d_2016, heagy_direct_2019}. Notably the hierarchical finite element approach developed in \cite{Weiss2017} enables complicated scenarios such as multiple lateral wells to be simulated. As compared to DC resistivity, time-varying EM experiments can be advantageous because they enable us to collect more data with the same survey geometry. The literature on inductive-source EM experiments in settings with steel-cased wells is relatively mature, with multiple authors having included magnetic permeability in their analyses \citep{augustin_theoretical_1989, Wu1994, kaufman_influence_1996, Wilt1996}. Recently, there has been growing interest in grounded-source EM experiments in settings with steel infrastructure. Grounded sources generate both galvanic and induced excitations of the subsurface.

For numerical simulations of EM in settings with steel casing, there have also been developments for finite volume or finite element simulations \citep{Um2015, commer_transient-electromagnetic_2015, haber_modeling_2016, heagy_modeling_2019}. Several authors have taken the approach of replacing a casing with a series of electric dipoles as supported by the analysis in \citep{cuevas_analytical_2014}, or adopted the related method-of-moments approach for simulating conductive infrastructure \citep{tang_three-dimensional_2015, patzer_steel-cased_2017, kohnke_method_2018, orujov_electromagnetic_2020}.

Although much work has been carried out on the use of EM and steel-cased wells, most of the efforts have focussed on conductivity only. This is reasonable since it is the high conductivity of steel, $\sim 10^6$ S/m, that primarily controls the response. Notably, \cite{cuevas_effect_2018} considers magnetic permeability in simulations of borehole to surface simulations in the frequency domain. They conclude that the effects of permeability, while real, are often not significant, but we do not know if those assessments carry over to other geometries and choices of physical properties.

For many applications of interest, we are faced with scenarios where the EM signals of targets are small. As instrumentation, and the sensitivity of field instruments, improves, so does our ability to detect subtle signals. This motivates scrutiny of the details of the components of an EM response. Although in principle, it is appreciated that the magnetic permeability of steel may affect the data, without carrying through with the simulations, we don't know how large the effect will be and whether it can be ignored, or if, by accounting for it, we can obtain more information from the EM data.

For more general applications of EM, there have been other studies on the impacts of magnetic permeability in EM experiments. Foundational literature on solutions for the response of a conductive, permeable sphere is developed by \cite{wait_conducting_1951, wait_conducting_1953, ward_unique_1959}. \cite{zhang_simultaneous_1999} demonstrates how magnetic permeability impacts 1D inversions of EM data. \cite{Pavlov2001} provides an analysis about the role of magnetic permeability in time-domain EM experiments. They rewrite the Maxwell-Ampere equations to show two implications of magnetic permeability: its effect on the  magnetization and its impact on the induction currents. They show the effect of permeability is to amplify a time-domain EM signal and also delay the decay in time. \cite{Noh2016} examine the impacts of permeability in a frequency domain EM. They perform numerical simulations and use the integral form of Maxwell's equations to gain insights about the role of permeability. They describe the impacts in terms of induction, magnetization, and coupling effects. Both of these papers consider inductive sources and use models of compact targets with moderate conductivities. The contribution of our work is to extend the analysis on the role of permeability to grounded source experiments with steel-cased wells that introduce extreme contrasts in both conductivity and permeability.

\begin{figure}
    \begin{center}
    \includegraphics[width=\textwidth]{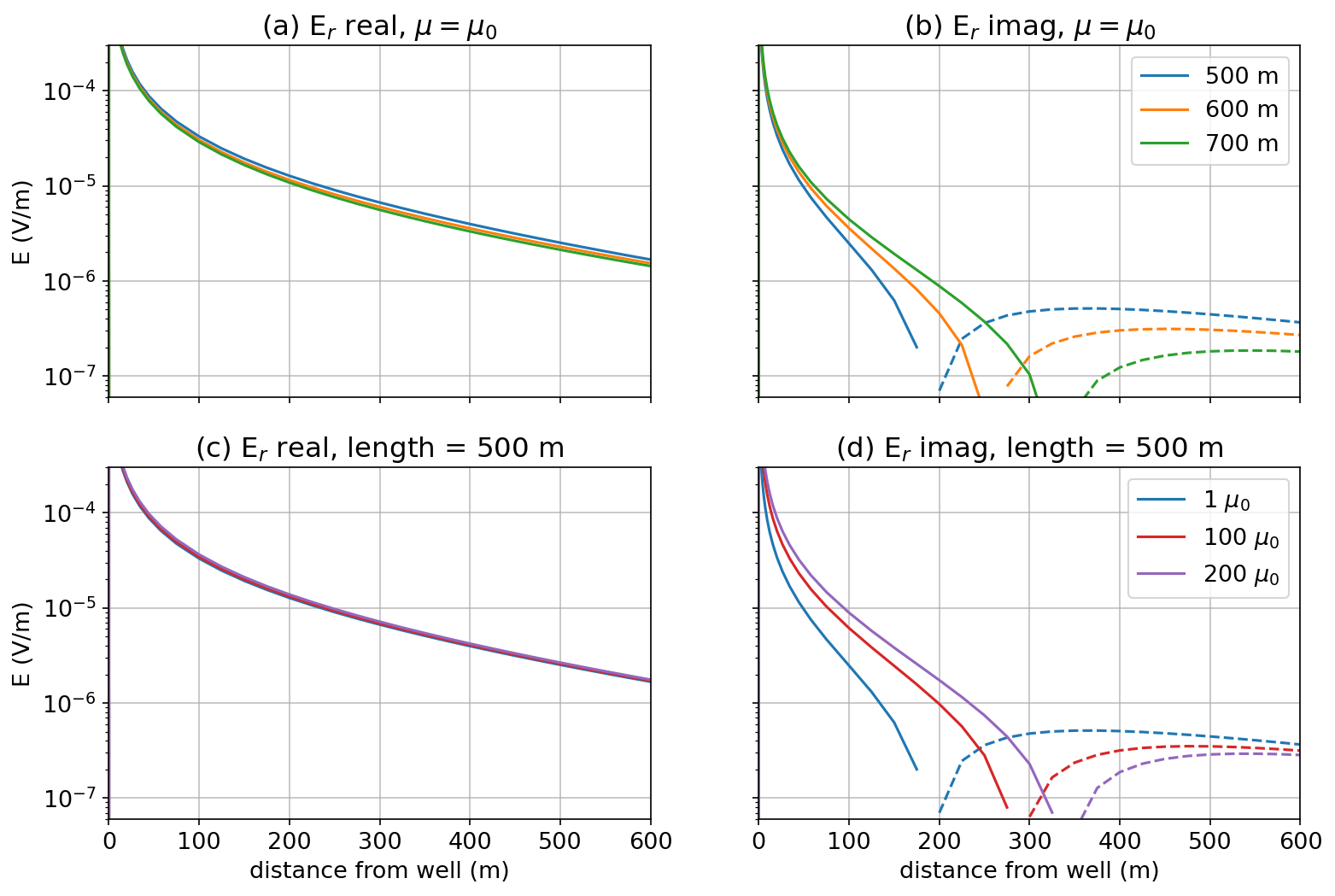}
    \end{center}
\caption{
    Radial electric field data for a top-casing experiment at 5 Hz. The top row shows the (a) real and (b) imaginary components for different lengths of a conductive well ($\sigma=5\times10^6$ S/m, $\mu=\mu_0$) in a 10$\Omega$m background. The bottom row shows the (c) real and (d) imaginary components for a 500m long well as the permeability is varied. The outer diameter of the casing is 10cm, and it is 2cm thick. The return electrode is 500m from the casing, and the receivers are along a line $180^\circ$ from the transmitter wire.
}
\label{fig:motivation-casing-integrity}
\end{figure}

As an illustrative example to help motivate our investigation, we consider the recent work by \cite{wilt_casing_2020}, which shows how surface EM fields from a top-casing electrode can be used for evaluating casing integrity. The E-field data are sensitive to the conductivity of the pipe, its length, and the electrical conductivity of the background. They showed how the amplitude of the response varies with the length of the pipe, and interestingly, that the imaginary part of the E-field response exhibited a change of sign along a profile extending away from the well. The distance from the well at which the cross-over occurs is correlated with the length of the well. It is also a function of the frequency at which the experiment is done and the conductivity of the background. Nevertheless, assuming that all of these variables are known, the location of the cross-over in the imaginary component could be a valuable first-order way to estimate the length of the pipe that was not broken. This is illustrated in Figure \ref{fig:motivation-casing-integrity}b, where we show the imaginary component of the radial electric field measured in a top-casing experiment for different lengths of a well. The casing has a conductivity of $5\times 10^6$  S/m, the background is $10  \Omega$m, and a 5Hz transmitter frequency is used. In this example, a cross-over noted at $\sim$325 m would correspond to a pipe of length L=700 m, whereas a 500m long well would have a crossover location near $\sim$175m.

If, however, we consider a fixed well length of 500m and instead vary the magnetic permeability, this also has a substantial impact on the imaginary component of the electric field. Figure \ref{fig:motivation-casing-integrity}d shows that the length of the pipe, inferred from a cross-over distance, can be significantly altered by permeability effects of the casing. The cross-over of the imaginary component for a 500m well with a permeability of 200$\mu_0$ is at a very similar location to a 700m well with a permeability of $\mu_0$. Although 5Hz is generally considered to be low-frequency, permeability has a non-trivial effect on the data. The detailed understanding of the complex behaviour of the EM fields deserves further attention and is one of the motivating factors for this paper.

The goal of this paper is to gain an understanding of the interplay between the EM fields in conductive and permeable materials. Additional current geometries are introduced because of the permeability and these are, at least for us, not intuitive. We focus on simple survey and casing geometries and examine the impacts of magnetic permeability in the EM response. Most of the analyses in the literature make use of simulations in the frequency domain. We find it insightful to view currents and fields in the time domain and use them to help understand what is happening in the frequency domain. In the end, simulating results in both frequency and time are mutually informative.

The paper is structured as follows. We begin, in Section 2,  by introducing the motivating example and using numerical simulations to illustrate how magnetic permeability impacts EM data in both the frequency and time domains. Re-writing Maxwell's equations to separate the magnetization and induction effects, as inspired by \cite{Pavlov2001, kaufman_principles_2009, Noh2016}, greatly aids in understanding the role of magnetic permeability. This section establishes how magnetic permeability impacts data and  motivates the subsequent sections which delve into the underlying physical mechanisms. The remainder of the paper seeks to develop the understanding of how permeability influences the EM response. In Section \ref{sec:tdem}, we dissect the time-domain EM response of a conductive, permeable casing. We show that the combination of the inductive and magnetization components of the response produce a poloidal current system within the casing and generate anomalous currents in the surrounding formation. In practice, many of the EM experiments conducted in settings with infrastructure are frequency domain experiments. Thus, in Section \ref{sec:fdem} we discuss how insights gained from the time-domain can be used to aid in understanding the frequency domain EM response. We finish with discussion on implications for the development of 3D numerical simulations and future advancement of the use of EM in settings with steel infrastructure.

\section{Impacts of permeability on EM data}
\label{sec:setup}
\subsection{The role of magnetic permeability in Maxwell's equations}
We begin by examining where magnetic permeability plays a role in Maxwell's equations. As noted by \cite{Ward1988, Pavlov2001, Noh2016} and others (e.g. \cite{revil_self-potential_2013} in the context of conductivity and self-potential), we can rewrite the quasi-static Maxwell's equations in time as
\begin{equation}
    \nabla \times (\nabla \times \vec{e}) - \nabla \ln \mu_r \times (\nabla \times \vec{e}) + \mu \sigma \frac{\partial \vec{e}}{\partial t} = -\mu \frac{\partial \vec{j}_s}{\partial t}
\label{eq:permeability-tdem}
\end{equation}

where $\vec{e}$ is the electric field, $\vec{j}_s$ is the source current density, $\mu_r$ is the relative permeability, and $\sigma$ is conductivity. This formulation follows from \citep{Pavlov2001}. There are two places where we see permeability playing an explicit role: the second term shows a dependence of the response on where there are changes in the permeability, and the third term shows a dependence of the response on the product of conductivity and permeability.

We can similarly write this equation in the frequency domain as
\begin{equation}
    \nabla \times (\nabla \times \vec{E}) - \nabla \ln \mu_r \times (\nabla \times \vec{E}) + i \omega \mu \sigma \vec{E} = -i\omega \mu \vec{J}_s
\label{eq:permeability-fdem}
\end{equation}

where here we use capital letters to denote frequency-domain fields and fluxes. Note that an $e^{i\omega t}$ Fourier transform convention is used.

In the remainder of this section, we use numerical simulations to illustrate how magnetic permeability influences both frequency and time domain EM data. This provides the motivation for the subsequent sections which delve into the details of the physical response of a conductive, permeable well in a grounded-source EM experiment.
\subsection{Setup}

\begin{figure}
    \begin{center}
    \includegraphics[width=0.8\textwidth]{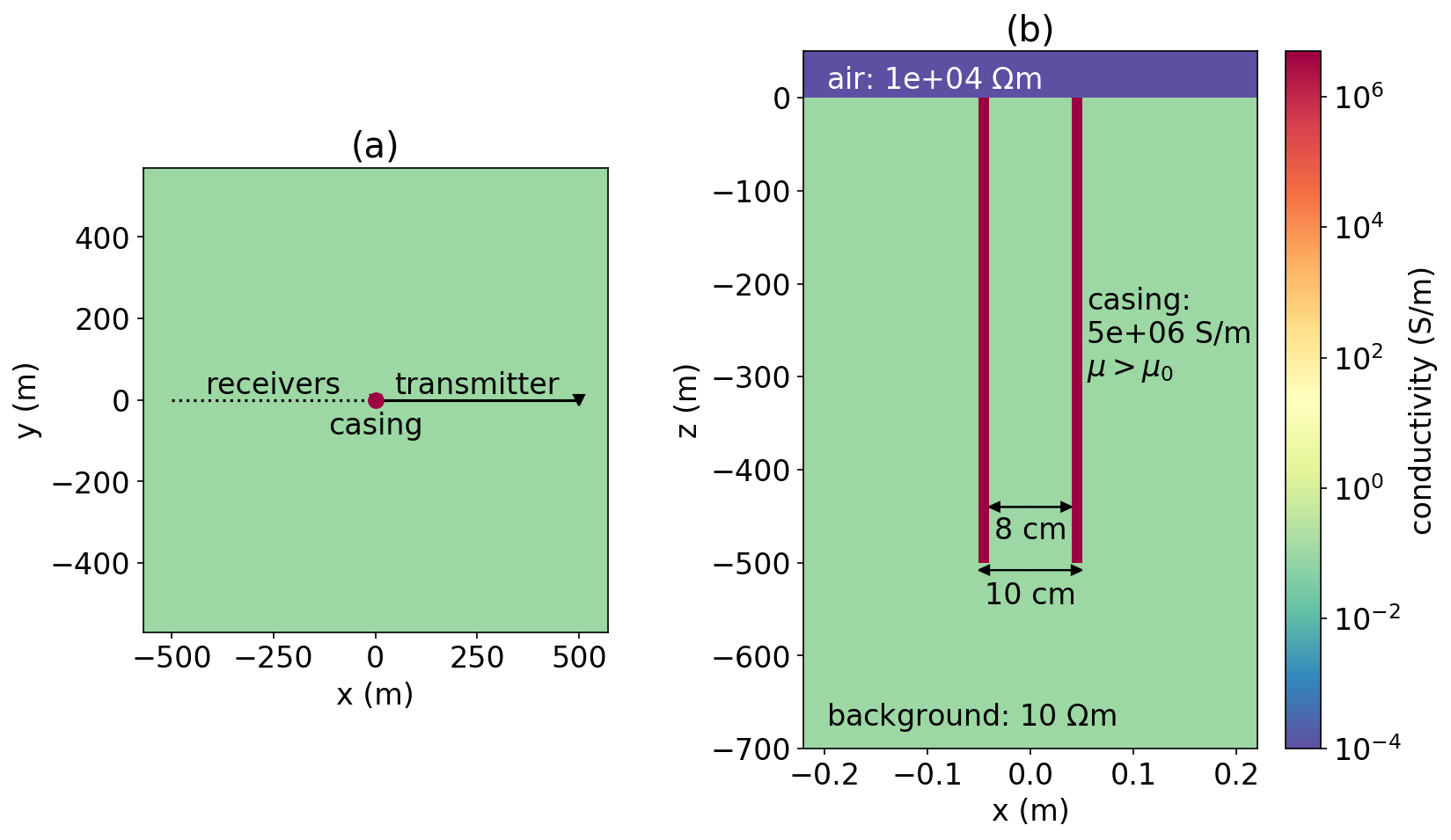}
    \end{center}
\caption{(a) Survey geometry and (b) geometry of the casing that we consider in this paper. A 500m long well with a conductivity of $5\times10^6$ S/m is positioned in a 10 $\Omega$ m halfspace. Data are collected along a line that is opposite to the transmitter wire.
}
\label{fig:setup}
\end{figure}

Our synthetic example, shown in Figure \ref{fig:setup}, is inspired by the CaMI site in Alberta which has been used as a study-site by multiple authors \citep{wilt_casing_2020, beskardes_effects_2021}. The well is 500m long and has a conductivity of $5\times10^6$ S/m. We will consider a range of permeabilities, noting that \citep{Wu1994} estimated values of $\mu_r \in [50, 150]$ for steel casing in lab experiments. The background conductivity is 0.1 S/m, which is similar to the background at the CaMI site \citep{um_joint_2020, lawton_development_2019}. The basic experiment we consider is a grounded source experiment where one electrode is connected to the top of the casing and the return electrode is 500m from the well.

For numerical simulations, we use the 3D cylindrical code described in \citep{heagy_modeling_2019} that discretizes in the vertical, radial, and azimuthal directions. In that paper we discussed validation of the code, including comparing numerical simulations for a time-domain EM experiment with those from \cite{commer_transient-electromagnetic_2015, Haber2007}. The mesh we use here discretizes the width of the casing with four 25mm cells and expands in thickness away from the well. We discretize azimuthally with 12 evenly-spaced cells. In the vertical direction, we use 5m cells in the region containing the well and expand outside of that. Padding cells in the radial and vertical directions ensure that the mesh extends beyond the skin depth for the frequency we are interested in or for the latest time in a time domain survey\footnote{Note that the value that we use for the resistivity of the air is more conductive than is typically used in EM simulations and inversions (often taken to be $\sim10^8 \Omega$m). Since the casing is so highly conductive, numerical errors can be introduced when there are extreme contrasts at the air-earth interface. In our experience, we have found that a contrast of $\sim10^{10} \Omega$m results in a numerically stable solution where the air is still sufficiently resistive as not to introduce artefacts in the simulated data. This is comparable to the values chosen for the resistivity of air in \citep{commer_transient-electromagnetic_2015, wilt_casing_2020}.}.

\subsection{Frequency domain data}
\begin{figure}
    \begin{center}
    \includegraphics[width=\textwidth]{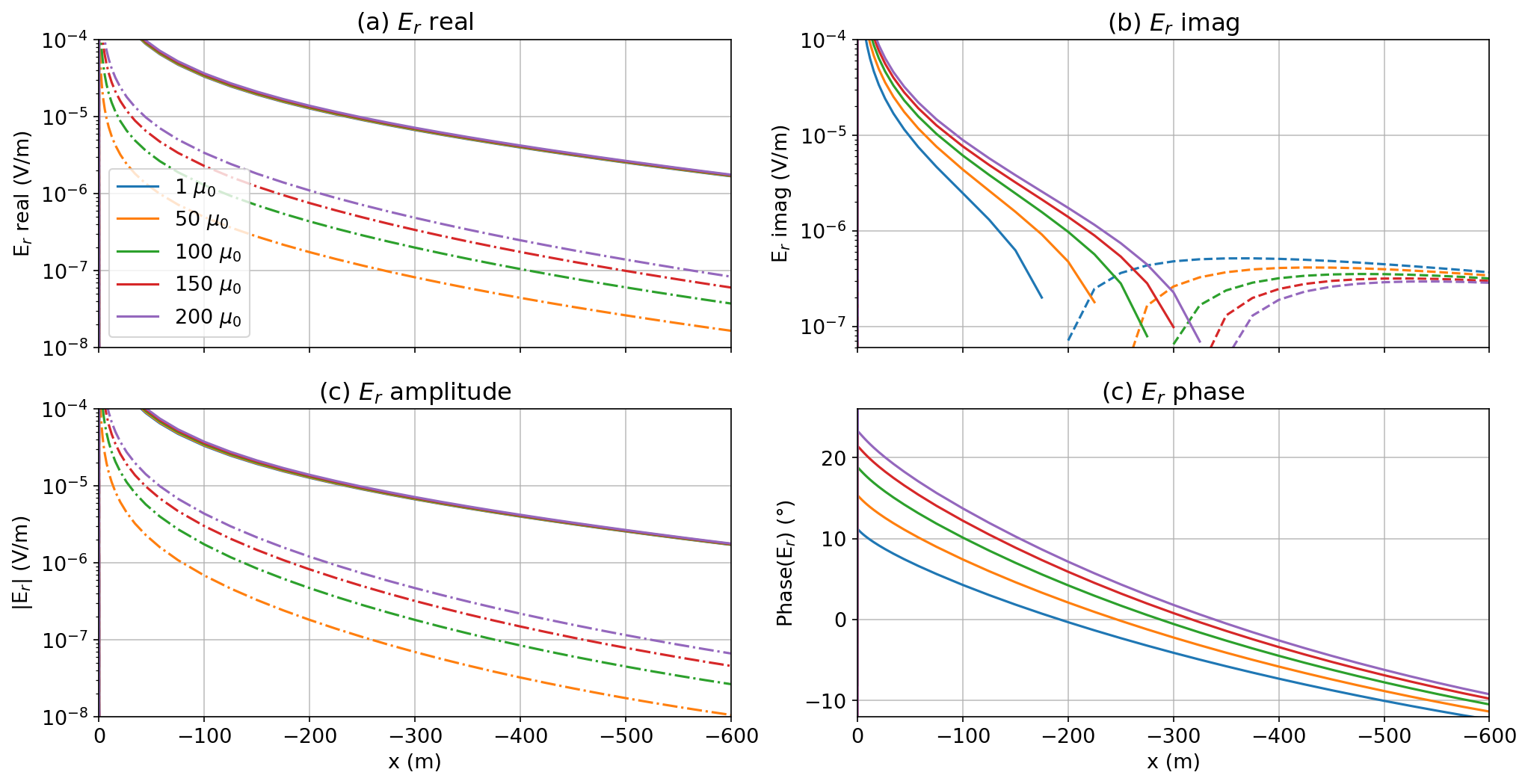}
    \end{center}
\caption{
    Radial electric field data for a top-casing experiment at 5Hz with the setup as shown in Figure \ref{fig:setup}. The top row shows (a) the real and (b) the imaginary components and the bottom row shows (c) the amplitude and (d) the phase. Solid lines indicate positive values (pointing away from the well) and dashed lines indicate negative values (pointing towards the well). In (a) and (c) we also show the difference between each of the permeable well scenarios and a non-permeable well ($\mu_r=1$) with the dash-dot lines.
}
\label{fig:e-fields-fdem}
\end{figure}

The majority of literature that looks at EM in settings with steel-cased wells considers frequency domain EM experiments. To examine the impacts of magnetic permeability in an FDEM experiment, we run simulations that vary the permeability from $\mu_r=1$ to $\mu_r=200$. In Figure \ref{fig:e-fields-fdem}, we show theradial electric field measured along a line opposite to the transmitter wire (as shown in Figure \ref{fig:setup}a).

Magnetic permeability has a substantial impact on the imaginary component. Between a well with $\mu_r=1$ and $\mu_r=150$, the location of the cross-over has moved by $>$100m. We can also see the impact in the phase; there is a difference of $10^{\circ}$ near the well between the $\mu_r=1$  and $\mu_r=150$ wells. This is comparable to the differences noted by \cite{cuevas_effect_2018} in numerical experiments or borehole-to-surface EM. The difference in the real component is less dramatic, but for a well with $\mu_r=150$, there is a 7\% difference from the non-permeable well at small offsets from the well. Since the real component is larger in magnitude than the imaginary components, the amplitude of the electric field is dominated by the behaviour of the real component, and thus less impacted by permeability.

\begin{figure}
    \begin{center}
    \includegraphics[width=\textwidth]{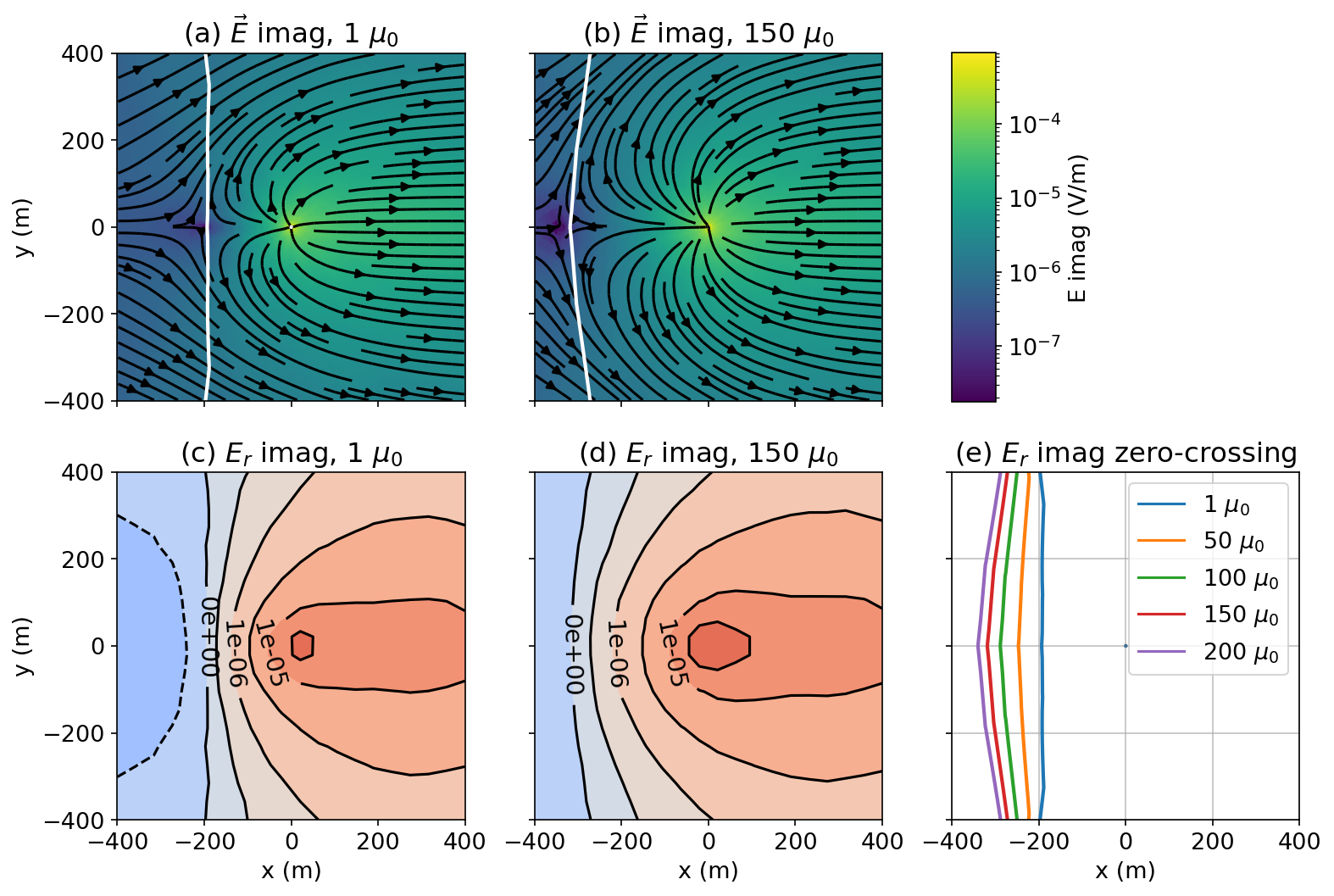}
    \end{center}
\caption{
    Depth slice of the imaginary component of the electric field just beneath the surface (z=-2.5m) for (a) a well with $\mu_r=1$ and (b) a well with $\mu_r=150$. The white line denotes the contour where the radial electric field is zero. In (c) and (d) we show contours of the radial electric field; red denotes positive values and blue are negative. In (e), we show the location of the zero-crossing for a range of casing permeabilities. The well is located at the origin (x=0m, y=0m). The transmitter wire runs from x=0m to x=500m along the y=0 line, as was shown in Figure \ref{fig:setup}a.
}
\label{fig:zero-crossing-permeability}
\end{figure}

Figure \ref{fig:e-fields-fdem} shows data along a single line opposite the transmitter wire, but a zero-crossing in the imaginary component can be observed for a range of azimuths on the opposite side to the transmitter wire. In Figure \ref{fig:zero-crossing-permeability}, we show a plan-view map of the imaginary component of the electric fields at the surface for a well with (a) $\mu_r=1$ and (b) $\mu_r=150$. The white line is the contour where the radial component of the electric field is 0 V/m. In (c) and (d), we have plotted contours of the radial component of the imaginary electric field, and in (e), we plot the contours for the zero crossing for different casing permeabilities. As noted by \cite{wilt_casing_2020}, increasing the frequency results in a larger amplitude of the imaginary component and pushes the zero-crossings out further from the well for frequencies up to $\sim$10 Hz. The location of the zero crossing in the imaginary component is sensitive to the properties of the casing, including the permeability.

To illustrate the impacts of permeability as a function of frequency, we choose the location $x=-100$m, $y=0$m and plot the radial electric field data measured at the surface for frequencies ranging from 0.1 Hz to 100 Hz. For this model, there is minimal impact on the real component for frequencies less than 2 Hz. As the frequency increases, we begin to see differences in the real component. At 10 Hz, which is typically considered ``low'' frequency, the real part differs by approximately 20\% between the model with a relative casing permeability of $\mu_r=150$ and a non-permeable well. The imaginary component is more substantially impacted by permeability; there is a factor of 4 between the data for $\mu_r=150$ and $\mu_r=1$ for low frequencies. These effects are also very evident in the amplitude and phase plots where phase differences of 5-10 degrees are evident. At shorter offsets, the magnitude of the fields, and the difference between the permeable and non-permeable well scenarios is larger. Similarly, with increasing distance from the well, the magnitudes and differences decrease.

\begin{figure}
    \begin{center}
    \includegraphics[width=0.8\textwidth]{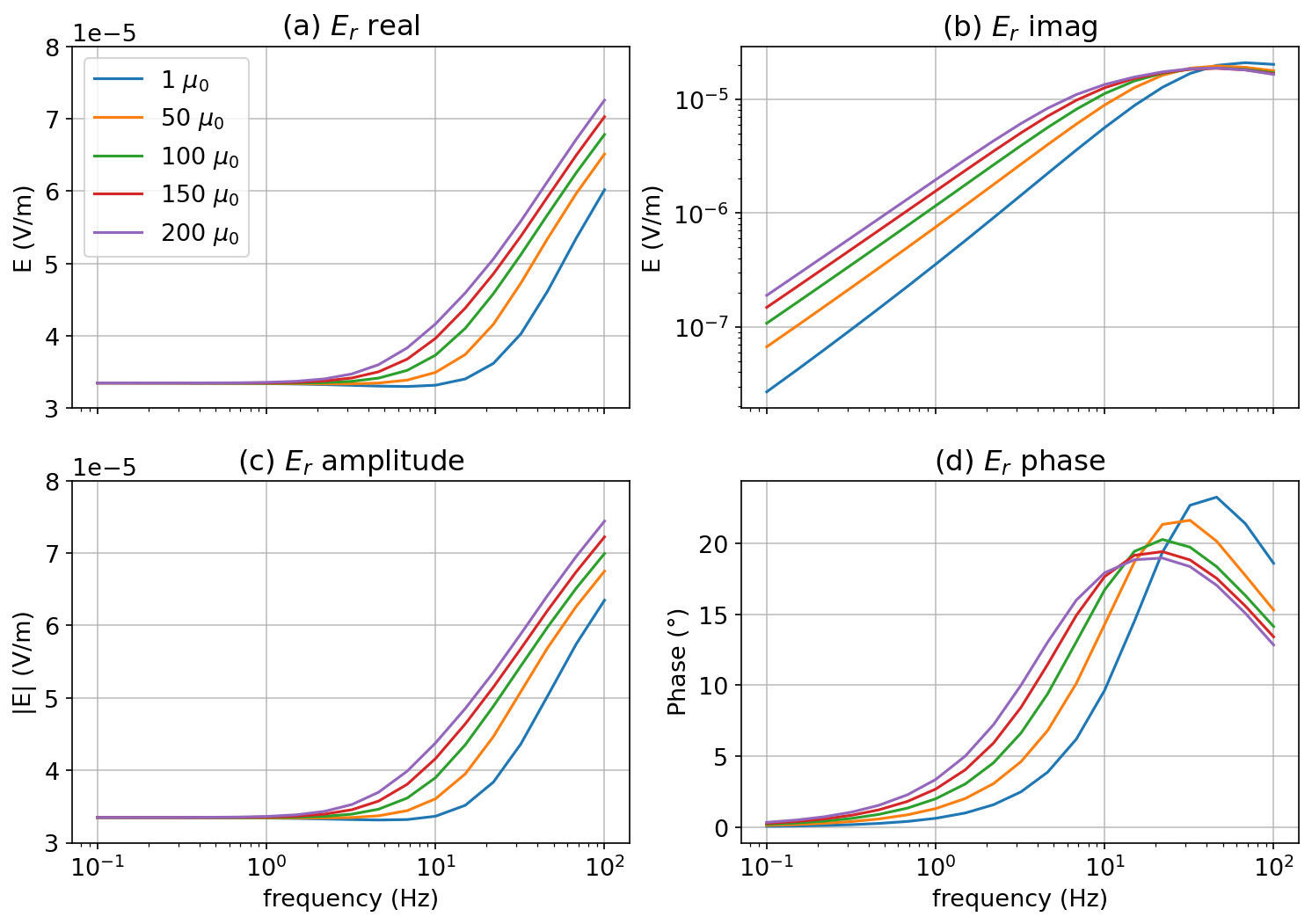}
    \end{center}
\caption{
    Radial electric field data at x=-100m, y=0m as a function of frequency.
}
\label{fig:data-100m-frequency}
\end{figure}

\subsection{Time domain data}
\begin{figure}
    \begin{center}
    \includegraphics[width=0.5\textwidth]{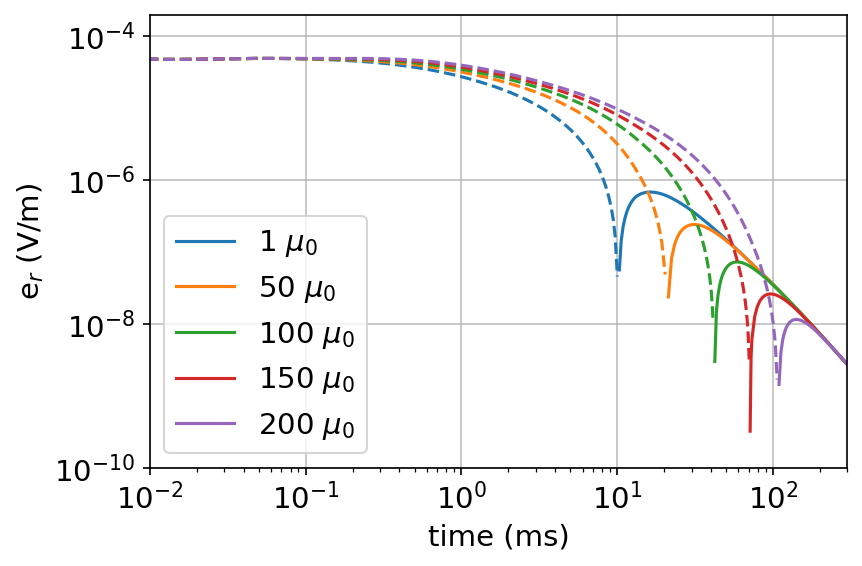}
    \end{center}
\caption{
    Radial electric field data at x=-100m, y=0m for a time domain EM experiment.
}
\label{fig:e-fields-tdem}
\end{figure}

Next, we examine the impacts of permeability in a time-domain experiment. In Figure \ref{fig:e-fields-tdem}, we show the radial electric field as a function of time at the location $x=-100$m, $y=0$m.  At times less than 1ms, there is minimal difference between the simulated data for each of the models. At 10ms, we can see a substantial difference between the permeable and non-permeable models. As has been noted by \citep{Pavlov2001} and others, permeability slows the decay, and similarly, the time at which we observe a sign-change in the radial component of the electric field. At sufficiently late times ($>$200ms), we no longer see the impacts of the casing in the data.

\bigskip

At this point, we have shown that permeability can have a non-negligible effect on both frequency and time domain EM data, but we have not yet discussed the physical mechanisms for how it is influencing the EM responses. The remainder of this paper is dedicated to understanding the role of permeability and the influence it has on EM responses. We will discuss why there is a zero-crossing in the frequency and time domain EM data, why it is sensitive to permeability, and illustrate how permeability alters current systems in an EM experiment.

\section{Time domain EM response of a conductive, permeable casing}
\label{sec:tdem}

\subsection{Currents in the formation}

To unravel the role of magnetic permeability and its impacts, we will examine the time-domain EM response of a conductive, permeable casing. Arguably, the behaviour of fields and fluxes through time is easier to reason about than a partition into real and imaginary components in a frequency domain experiment. So we will reason about the physics in the time-domain and relate our insights back to the frequency domain in the next section.

\begin{figure}
    \begin{center}
    \includegraphics[width=\textwidth]{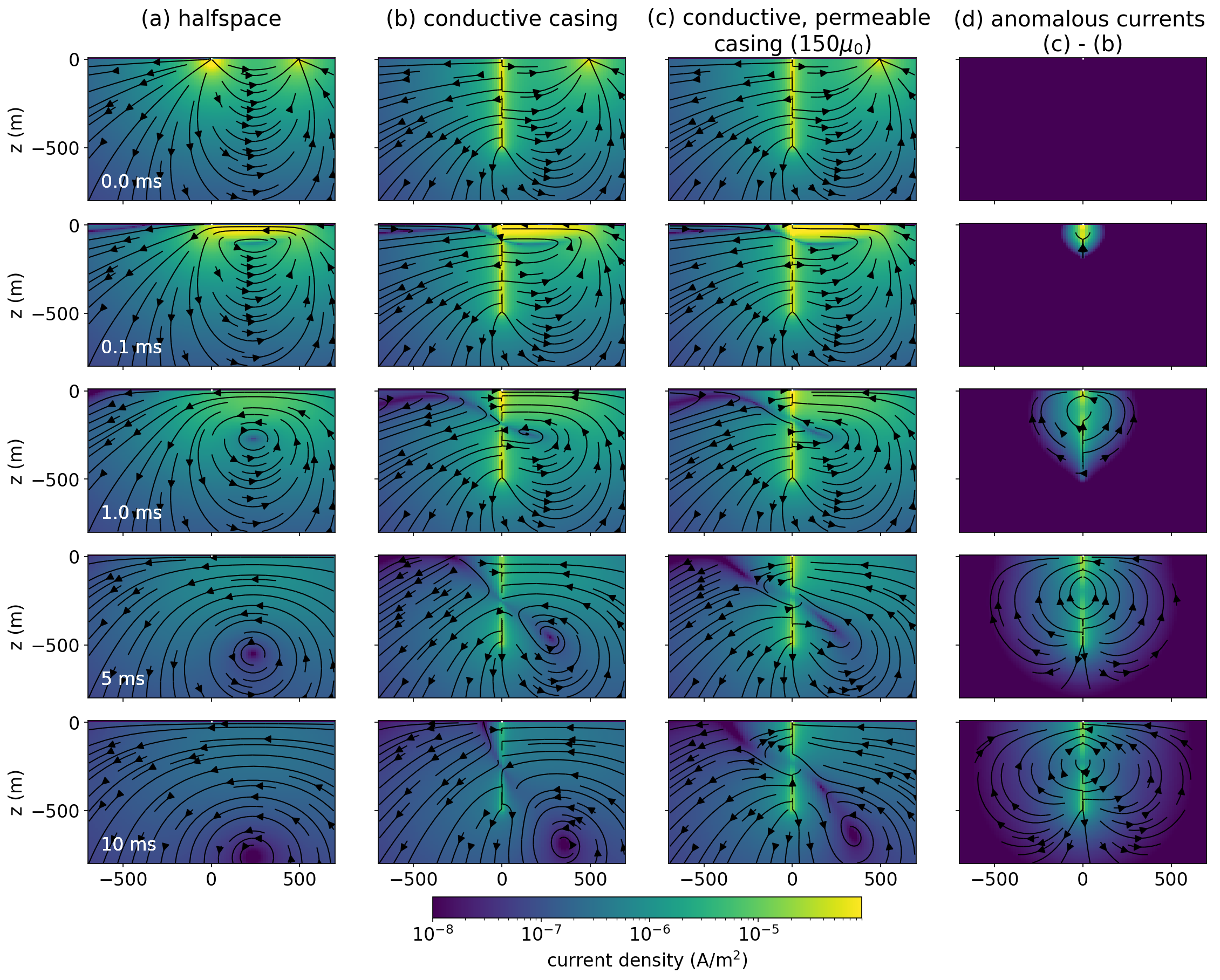}
    \end{center}
\caption{Cross sections showing the current density through time for a time domain EM experiment over (a) a half-space, (b) a conductive well ($5\times10^6$ S/m, $\mu_0$), and (c) a conductive, permeable well ($5\times10^6$ S/m, $150\mu_0$). The panel on the right shows (d) the difference between the conductive, permeable well and the conductive well. The threshold for the streamplot arrows is the same as the colorbar minimum ($10^{-8}$A/m$^2$)
}
\label{fig:tdem-cross-section-currents}
\end{figure}

We use the same setup as shown in Figure \ref{fig:setup}. For simplicity, we consider a step-off waveform. In Figure \ref{fig:tdem-cross-section-currents} we show a cross section of currents through the earth for 3 models: (a) a halfspace, (b) a halfspace that includes a conductive casing ($5\times10^6$ S/m), and (c) a halfspace with a casing that is conductive and permeable ($5\times10^6$ S/m, $150\mu_0$).

We start by examining the currents in the halfspace (Figure \ref{fig:tdem-cross-section-currents}a). The top row, at t=0ms, is the DC resistivity solution. After t=0, the current in the transmitter is shut off, and image currents, which oppose the change in magnetic field, are induced in the Earth \citep{nabighian_quasi-static_1979}. These currents are in the same direction of the current in the source wire, as to preserve the magnetic field, and this causes a circulation of current as the galvanic and image currents interact. Both currents diffuse down and out through time.

With a conductive casing (Figure \ref{fig:tdem-cross-section-currents}b), at t=0ms, charges are distributed along the length of the well. As a result, there are radial currents, sometimes referred to as ``leak-off'' currents emanating from the well. \cite{kaufman_transmission-line_1993} discuss the distribution of charges and currents for two limiting cases, a ``short'' well and a ``long'' well, where the length scales are determined by the product of the conductivity of the casing and its cross-sectional area, as well as the conductivity of the background\footnote{\cite{schenkel_electrical_1994} defines the ``conduction length'' of the casing as $\delta = \sqrt{\rho_f A_c \sigma_c}$ where $\rho_f$ is the resistivity of the formation, $A_c$ is the cross-sectional area of the casing, and $\sigma_c$ is the conductivity of the casing. A ``short'' well is one whose length is shorter than the conduction length and a ``long'' well has length larger than the conduction length. }. For short wells, charges are distributed approximately uniformly along the length of the well. For long wells, there is an exponential decay in the charge density and therefore leakage currents along the length of the well. In \cite{heagy_modeling_2019} we compare these limits to numerical simulations of a DC resistivity experiment.

\begin{figure}
    \begin{center}
    \includegraphics[width=\textwidth]{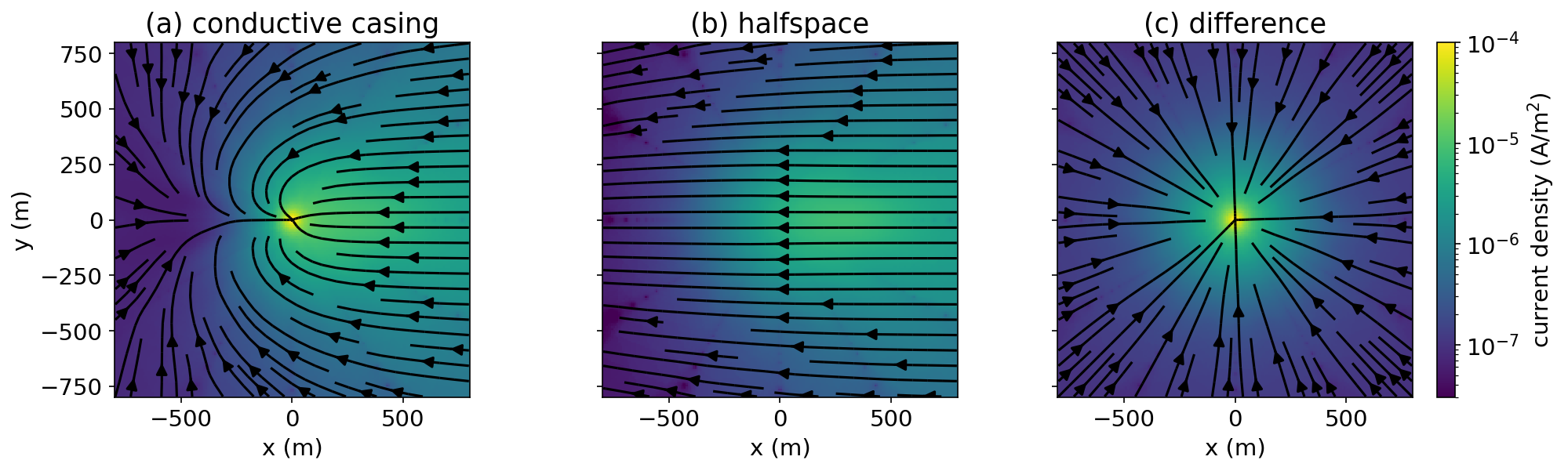}
    \end{center}
\caption{
    Depth slice at z=-2.5m and t=1ms showing the currents in an experiment with (a) a conductive casing ($\mu=\mu_0$),
    and (b) a halfspace. The difference between the two scenarios (casing - halfspace) is shown in (c).
}
\label{fig:currents-depth-slice}
\end{figure}

When the current in the transmitter is shut off, image currents are induced, but when a well is present, we have the added complication of current channeling into the coductive casing. This causes a ``shadow'' zone to develop on the side of the casing opposite the transmitter wire. Similar effects are seen in the frequency-domain, as illustrated by \cite{wilt_casing_2020}. This ``shadow-zone'' can be understood to be the interaction between the image currents and the currents that are channelled into the casing. In Figure \ref{fig:currents-depth-slice} we show a depth slice of the currents at t=0.1ms just below the surface (z=-2.5m). In (a), we show the simulated currents for a conductive casing (with $\mu_r = 1$), in (b) the currents in the half-space simulation, and in (c) the difference (conductive casing minus the halfspace). The image currents in the halfspace point towards the left whereas the secondary currents due to the casing (the channelled currents) point towards the well. The zero-crossing in the radial electric field corresponds to the location where the halfspace image currents are equal to and in opposite direction of the channelled currents.

If we now turn our attention to the conductive, permeable well, in Figure \ref{fig:tdem-cross-section-currents}(c), we see that at t=0ms, the currents are identical to the conductive well. There is no influence of magnetic permeability in the electrostatic / DC limit. However, the impacts of permeability are seen later in time as the currents decay more slowly in the casing and surrounding formation; this is consistent with the delay in the decays shown in Figure \ref{fig:e-fields-tdem}. In panel (d), we show the difference between the permeable and non-permeable casing scenarios. At early times, the largest difference broadly aligns in depth with where the image current is. At later times, the image current has diffused past the length of the well, and we see differences along the entire depth-extend of the well. We also note that the difference is cylindrically symmetric, having only radial and vertical components.

\subsection{Enhanced excitation with permeability}

\begin{figure}
    \begin{center}
    \includegraphics[width=\textwidth]{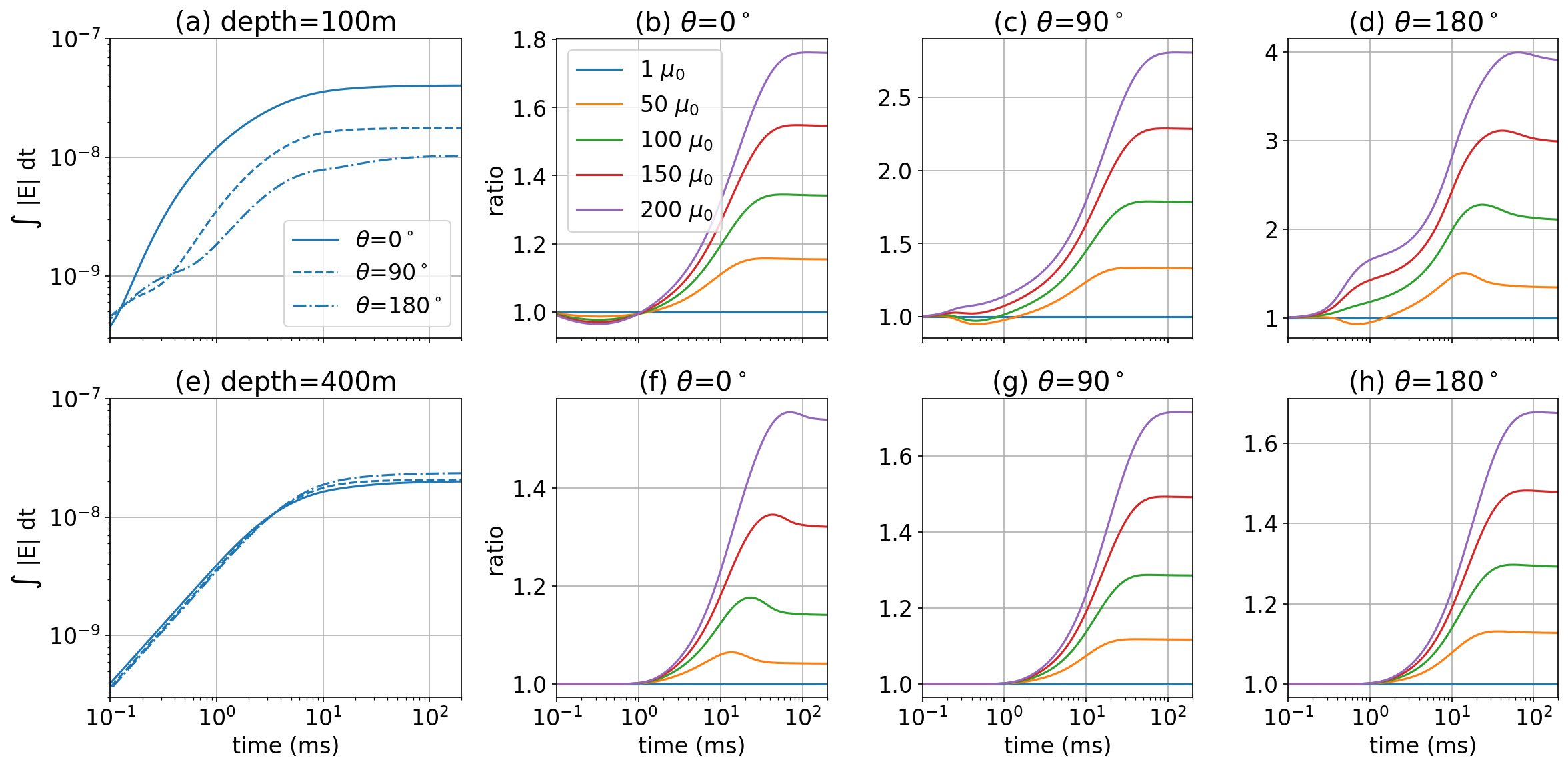}
    \end{center}
\caption{
    Integral of the average electric field in a test volume through time, which we use as a proxy for excitation. The test volume extends from 50m - 100m radially and 50m vertically centered about depths of (a) 100m and (e) 400m. The different line-styles in (a) and (e) indicate different azimuths ($0^\circ$ is under the transmitter wire, $90^\circ$ is orthogonal to it, and $180^\circ$ is opposite to it) for the non-permeable well ($\mu=\mu_0$). Panels (b), (c), and (d) show the ratio of the excitation with respect to the non-permeable well ($\mu=\mu_0$) for the test volume centered at 100m depth. Panels (f), (g) and (h) show the ratios of the excitation for the test volume centered at 400m depth.
}
\label{fig:excitation-time-integrated}
\end{figure}

It was noted in \cite{cuevas_effect_2018} that permeability may enhance the inductive component of the EM response, and in \cite{heagy_electrical_2022}, we showed that increased permeability can improve our ability to detecting targets with electric field data collected on the surface. To illustrate how detectibility might be enhanced we appeal to some basic principles. In the resistive limit, the strength of the anomalous currents depend upon the magnitude of the primary electric field \citep{west_physics_1991}. Thus, to first-order, if the magnitude of the primary electric field is doubled, the excitation of the target doubles. To quantify excitation in our example, we consider a test volume in the formation, compute the average electric field in this volume and integrate it as a function of time. The volume we choose extends radially from r=50m to r=100m and 50m vertically. We consider 3 different azimuths, inline with the transmitter wire ($\theta = 0^\circ$), orthogonal to the wire ($\theta = 90^\circ$), and opposite the wire ($\theta = 180^\circ$). Figure \ref{fig:excitation-time-integrated} shows the time-integrated electric field in this volume at (a) 100m depth, and (e) 400m depth, for the non-permeable well ($\mu=\mu_0$).

For a shallow target, the excitation is greatest underneath the transmitter wire ($\theta=0^\circ$), whereas for a deeper target, either side ($\theta=0^\circ$ or $\theta=180^\circ$) provides comparable excitations. In Figure \ref{fig:excitation-time-integrated}(b-d) and (f-h), we show the ratio of the average electric field for different casing permeabilities with the scenario where $\mu=\mu_0$. At early times, the ratio is 1 as permeability has minimal impact. At later times, permeability provides additional excitation. For a shallow target, a casing with a permeability of 150$\mu_0$ results in an excitation that is a factor of $>50\%$ larger as compared to a non-permeable well after 10ms. For a deeper target, the effect is smaller but not inconsequential, with the additional excitation ranging from 30\%-50\% for a casing with a permeability of 150$\mu_0$. Since the additional currents due to the permeability of the well, seen in Figure \ref{fig:tdem-cross-section-currents}(d) are cylindrically symmetric, the dependence of the additional excitation on azimuth is a result of the survey geometry.

\subsection{Currents within the casing}
\begin{figure}
    \begin{center}
    \includegraphics[width=1\textwidth]{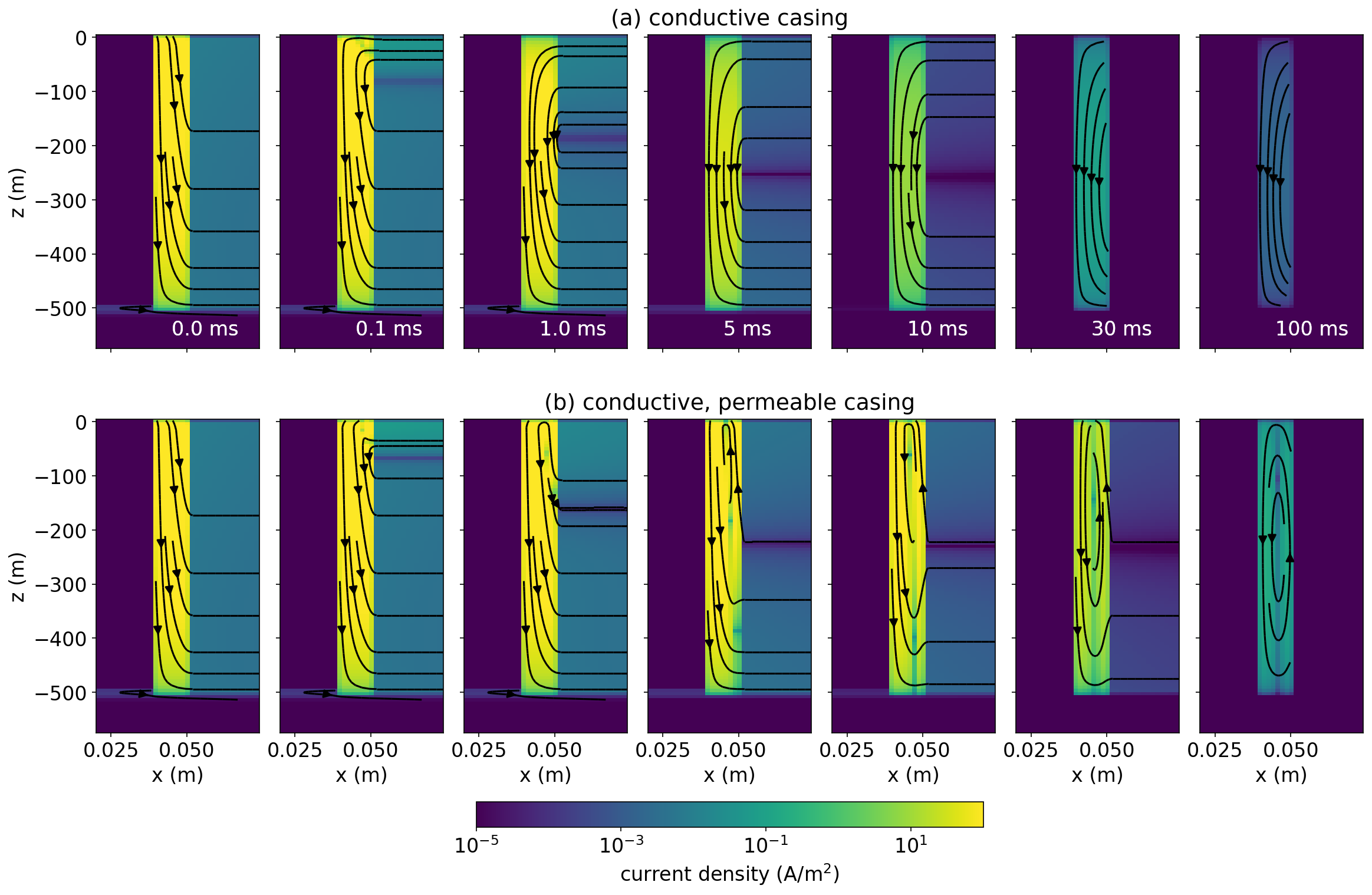}
    \end{center}
\caption{Cross sections of currents within (a) a conductive casing ($5\times10^6$ S/m, $\mu_0$) and (b) a conductive, permeable casing ($5\times10^6$ S/m, $150\mu_0$). Not to scale.
}
\label{fig:tdem-casing-currents}
\end{figure}

The additional currents arising as a result of permeability are not simply amplifying the currents due to a conductive casing, there is also an impact on the geometry of the currents, as can be seen in Figure \ref{fig:tdem-cross-section-currents}. To examine why this is, we zoom in to the currents within the casing in Figure \ref{fig:tdem-casing-currents}. The top row shows the conductive well and the bottom row shows the conductive, permeable well ($150\mu_0$). There are two main features to note. First, the magnitude of the currents (indicated by colour) is larger at later times in the permeable well than in the conductive well, particularly after $\sim$5ms. The other feature to note is the geometry of the currents. For the conductive well, we see that the currents are flowing downwards in the casing through time. However, for the conductive, permeable well, we see that at later times, a poloidal current system develops where currents flow downwards along the inner casing wall and upwards near the outer edge of the casing.

\begin{figure}
    \begin{center}
    \includegraphics[width=\textwidth]{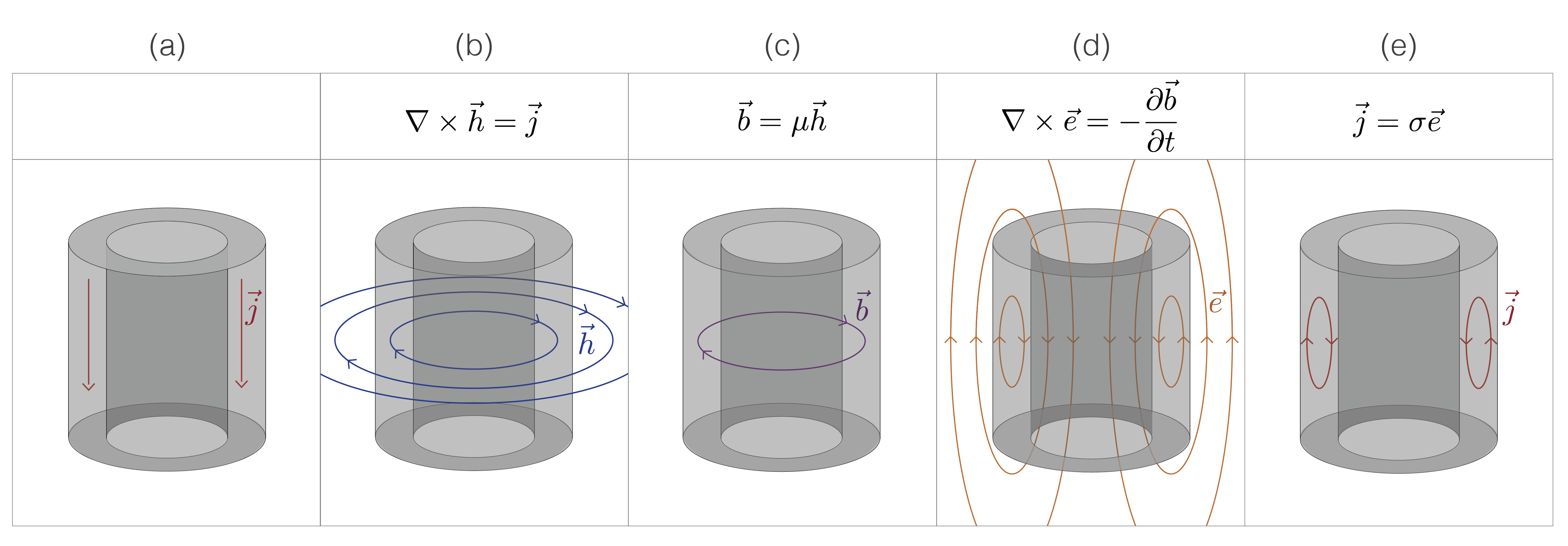}
    \end{center}
\caption{
    Sketch demonstrating how a poloidal current system can arise inside of a conductive, permeable casing.
    A source current is applied and (a) currents flow downwards through the pipe. (b) Currents generate rotational magnetic fields according
    to Ampere's law. (c) Magnetic flux density is concentrated in the permeable pipe according to the constitutive relation between $\vec{b}$ and $\vec{h}$.
    (d) The magnetic flux varies with time creating rotational electric fields according to Faraday's law.
    (e) Currents associated with those electric fields are concentrated in conductive regions of the model in accordance with Ohm's law.
}
\label{fig:casing-mu-sketch}
\end{figure}

To understand this poloidal current system, we refer to Maxwell's equations. We illustrate this in Figure \ref{fig:casing-mu-sketch}: (a) a current is applied to the casing; (b) by Ampere's law, vertical currents produce a toroidal magnetic field; (c) by the constitutive relationship between the magnetic flux density and the magnetic field (Ohm's law for magnetics), magnetic flux is concentrated inside of permeable materials; (d) the magnetic flux is changing through time which creates a poloidal electric field; (e) currents are concentrated in conductive materials according to Ohm's law leading to a poloidal current system.

\subsection{Explaining the poloidal current system}
The cartoon in Figure \ref{fig:casing-mu-sketch} is obviously a simplification of the physics, but it provides a useful conceptual model. To approach understanding this poloidal current system in a more quantitative manner, we return to equation \ref{eq:permeability-tdem}. Using $\nabla \times \vec{e} = -\partial\vec{b}/\partial t$ we rewrite equation \ref{eq:permeability-tdem} as
\begin{equation}
    \frac{\partial}{\partial t} \left(
        \nabla \times \vec{b}
        - \nabla \ln \mu_r \times \vec{b}
        - \mu \sigma \vec{e}
        - \mu \vec{j}_s
    \right)
        = 0
\label{eq:permeability-tdem-dbdt}
\end{equation}

This must hold for all times, and using Ampere's law, it can be shown that the term under the time-derivative is equal to zero. Thus, we have
\begin{equation}
\nabla \times \vec{b} - \nabla \ln \mu_r \times \vec{b} - \mu\sigma\vec{e} = \mu\vec{j}_s
\label{eq:permeability-ampere}
\end{equation}

Away from the source, $\vec{j}_s=0$, giving
\begin{equation}
\nabla \times \vec{b} = \nabla \ln \mu_r \times \vec{b} + \mu\sigma\vec{e}
\label{eq:permeability-ampere-no-source}
\end{equation}

Both terms on the right hand side explicitly contain the permeability. Following \cite{Pavlov2001, Noh2016}, we refer to these as the magnetization and induction terms, respectively. For the magnetization term, we note that $\nabla \ln \mu_r$ is zero everywhere except where there is a discontinuity in the relative permeability; in our problem this occurs at the walls of the casing.

\begin{figure}
    \begin{center}
    \includegraphics[width=\textwidth]{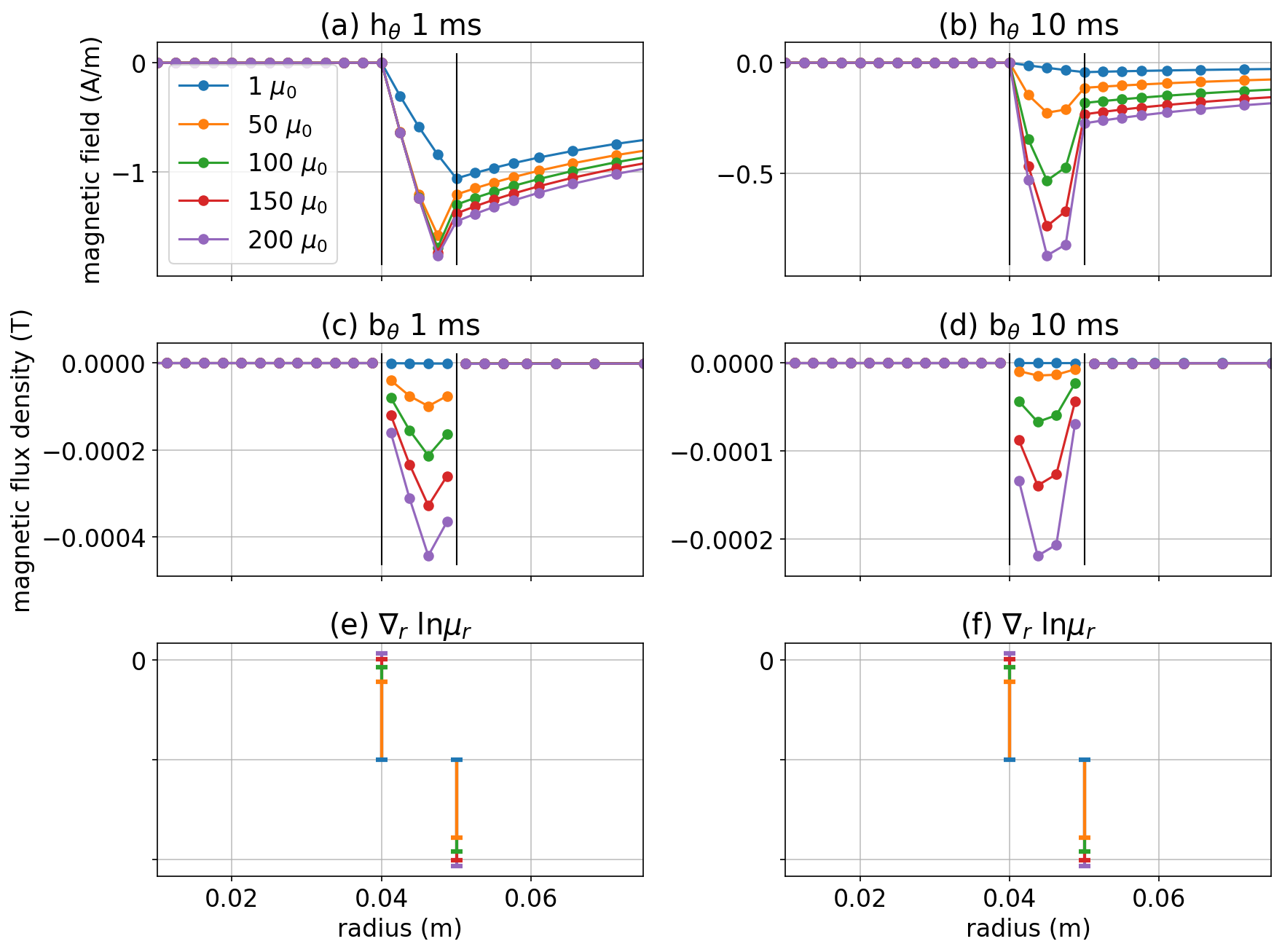}
    \end{center}
\caption{
    (a-b) Azimuthal component of $\vec{h}$ at 1ms and 10ms. (c-d) Azimuthal component of $\vec{b}$. (e-f) Radial component of $\nabla \ln \mu_r$. All plots are at 100m depth.
}
\label{fig:magnetization-term}
\end{figure}

To examine what the magnetization term contributes to the EM response, we choose a depth of 100m and plot $\vec{h}$, $\vec{b}$ and $\nabla \ln \mu_r$ in the vicinity of the casing in Figure \ref{fig:magnetization-term}. Due to the symmetry of the problem, the radial and vertical components of the magnetic field are negligible and we focus on the azimuthal component. The top row shows $h_\theta$, which is continuous across the casing walls, and the second row shows $b_\theta$ which is discontinuous where there is a discontinuity in permeability. The divergence of the log-permeability gives us two delta functions when $\mu_r > 1$, a positive value on the inside of the casing and a negative value on the outside. Since there is negligible current in the hollow interior of the well as compared to the casing, the current enclosed by an amperian loop with radius equal to that of the inner casing wall is negligible. Thus $\vec{b}$ on the inner casing wall is negligible. Therefore, the main contribution that the magnetization term makes to the response is on the outer casing wall. Since $\nabla_r \ln \mu_r$ is negative at the outer casing wall and $b_\theta$ is too, their cross product is a positive quantity in the z-direction. We can interpret this as a current that is scaled by the permeability, as shown on the right hand side of equation \ref{eq:permeability-ampere}. This means the magnetization term contributes an upward current on the outside portion of the casing.

\section{Frequency domain EM response of a conductive, permeable well}
\label{sec:fdem}

In practice, many of the EM experiments that are conducted in settings with steel infrastructure are frequency domain EM experiments. Thus, it is important to discuss how the frequency domain EM response is affected by magnetic permeability.
In Appendix \ref{app:fdem}, we repeat the analysis performed in Section \ref{sec:tdem}, but in the frequency domain. In this section, we will discuss the key take-aways and refer interested readers to the Appendix for further discussion of the details.

When moving from the time domain to the frequency domain, we can translate our understanding of how permeability influences the EM response by recognizing that responses at early times are analogous to high frequencies and those at late times are analogous to low frequencies. An important distinction is that, in a frequency domain experiment, the transmitter is always on, meaning the real component always contains a galvanic or DC component.

In the time domain, the permeability of the casing made a larger impact at later times, implying we should observe impacts of permeability at low frequencies. At low frequencies, permeability can have a substantial impact on the imaginary components, but whether it impacts the real component depends on the magnitude of the inductive response relative to the galvanic response. For example, in Figure \ref{fig:data-100m-frequency}, where we showed electric field data on the surface as a function of frequency, we see that permeability has no noticeable impact on the real component below 1 Hz for this example. In the imaginary component, there is a factor of 3 difference between the well with $\mu_r = 150$ and a non-permeable well, but this translates to a phase difference of only a few degrees because the galvanic component is so dominant. As the frequency is increased, the inductive part of the response also increases. At 10 Hz, the inductive component of the response is substantial, and for the example in Figure \ref{fig:data-100m-frequency}, the real part of the electric field for the well with $\mu_r = 150$ is 20\% larger than the non-permeable well. Similarly there is a $>$20\% difference in the amplitude.

The zero-crossing observed in the imaginary component of data taken in frequency domain surveys, and used as a potential diagnostic for well length integrity \citep{wilt_casing_2020}, can be understood by referring back to the time domain. The zero-crossing occurs when the currents channelled into the well exactly oppose the image currents. Increasing the permeability of the casing increases current channelling into the casing (Figure \ref{fig:tdem-cross-section-currents}d), so increasing the permeability shifts the zero crossing in the imaginary component further away. We do not observe a zero-crossing in the real component because the real component is dominated by the galvanic currents.

In the previous section, we demonstrated how the anomalous currents due to increased permeability of a well may improve our ability to excite a target, particularly at later times (Figure \ref{fig:excitation-time-integrated}). With frequency, we observe similar anomalous currents, but since the transmitter is always on, whether these additional currents translate to additional excitation depends upon the frequency and the position of the target of interest with respect to the well.

\section{Discussion}
We have shown that permeability influences the EM response of a grounded-source experiment with steel cased wells in two ways:
\begin{enumerate}
\item it enhances the induction component of the response
\item it introduces a magnetization current on the outer casing wall that opposes the induction currents.
\end{enumerate}
Faraday's law couples the induction and magnetization components in a time or frequency domain experiment and, as a result, a poloidal current system develops within the casing. For the current system to arise, the casing must be both highly conductive and magnetic.

\begin{figure}
    \begin{center}
    \includegraphics[width=0.6\textwidth]{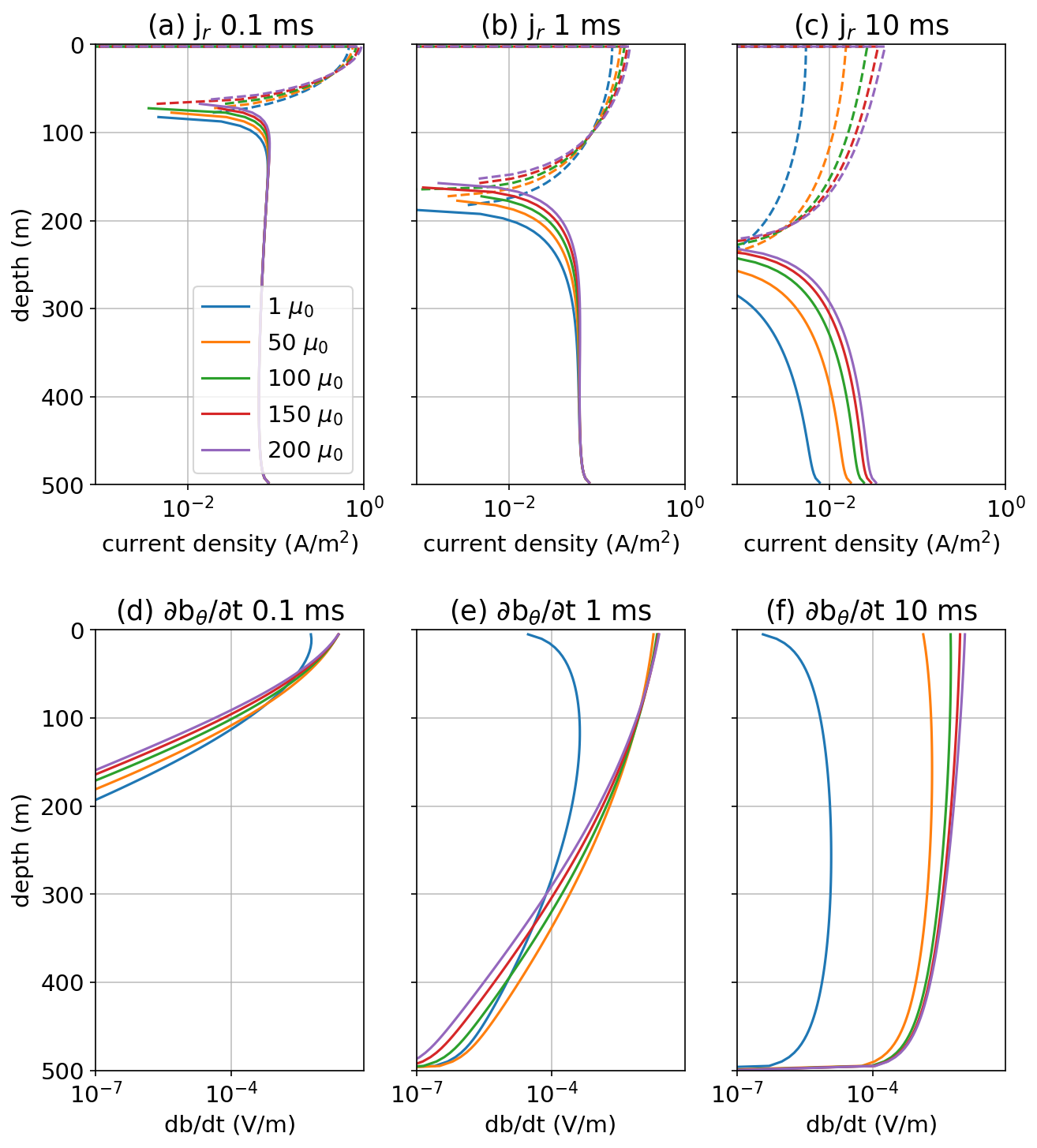}
    \end{center}
\caption{
    (a-c) Radial component of the currents at the outer radius of the casing (also referred to as the ``leak-off'' currents) in a time-domain experiment.
    (d-f) Azimuthal component of $\partial \vec{b}/\partial t$ within the casing.
}
\label{fig:permeability-contributions}
\end{figure}

The two impacts of permeability on the currents within the casing translate to implications in the surrounding formation: (a) additional radial ``leak-off'' currents change the radial component of the current density within the formation, and (b) the amplification of the azimuthal component of $\partial \vec{b}/\partial t$ within the casing can alter the radial and vertical currents in the formation. We have shown both of these contributions in Figure \ref{fig:permeability-contributions}; (a-c) show the radial current density at the outer casing wall at three different times, and (d-f) show $\partial b_\theta / \partial t$ within the well at these same times. The contribution of these two terms is responsible for the anomalous currents within the formation that are observed in Figure \ref{fig:tdem-cross-section-currents}.

The anomalous currents can affect the excitation of a target within the formation. Moreover, in a time-domain experiment, an increased permeability slows the decay of currents in the well and provides a longer time-window over which a target may be excited. This could be advantageous for helping detect a target in a time-domain EM experiment. In a frequency domain experiment, the source field is alway on, and therefore whether the anomalous currents enhance or reduce our ability to excite a target depends upon the frequency and the location of the target.  As we showed, even in experiments that would generally be considered ``low frequency'' (e.g. $<$ 10 Hz), permeability can have a measurable impact on data collected at the surface.

The role of permeability in the EM response also has implications for how simulations involving conductive, permeable casings can be achieved numerically. On a practical note, when discretizing the casing with standard finite volume or finite element codes, the mesh must be fine-enough in the radial direction in order to be able to simulate a poloidal current. This could not be accomplished if the mesh was only a single cell wide. By using a cylindrical mesh, we are able to sufficiently refine the mesh without enormous computational cost. However, a cylindrical mesh is limited in the geometries that can be simulated. Horizontal or deviated wellbore geometries cannot be captured with a cylindrically symmetric mesh. To simulate more complex, 3D scenarios, multiple authors have suggested replacing a conductive casing with a series of current elements or electric dipoles \citep{cuevas_analytical_2014}, or using a related method of moments approach \citep{kohnke_method_2018, tang_three-dimensional_2015} for simulations. Several authors have suggested that such approaches could be extended to include the impacts of permeability by including a model of magnetic dipoles along the axis of the casing (e.g. \cite{patzer_steel-cased_2017, kohnke_method_2018}). However, this would imply that the anomalous currents are in the azimuthal direction, which is not what we observe in a grounded-source EM experiment. How to capture the effects of permeability in a practical manner in 3D numerical simulations is an area for future research. A further complicating factor is that, in practice, the magnetic permeability of steel casings is generally unknown. Thus there are also research opportunities in the development of strategies for estimating casing properties from EM data.

\section{Conclusions}
In this paper, we have addressed the problem of understanding how magnetic permeability contributes to the EM response of a conductive, permeable well in grounded source EM experiments. As others have shown, variable magnetic permeability contributes to the EM response through magnetization and induction components. The interplay of these terms is particularly interesting in the context of steel-cased wells because steel is orders of magnitude more conductive than the surrounding geology.

Within the casing, the combination of the magnetization and induction terms results in a poloidal current system. The nature of this response is important for several reasons. First, the permeability of a well can alter the geometry and the magnitude of currents in the surrounding geologic formation. For certain survey geometries, this can be advantageous for exciting a response in a target of interest. Second, our results illustrate the potential importance of including permeability in numerical simulations of EM experiments in settings with steel infrastructure. This poses a practical challenge because standard finite volume or finite element approaches require that the mesh be refined sufficiently to capture the fine-scale effects within the casing, while being large enough to simulate the geologic structures of interest. We circumvented this challenge by working with a simple model of a vertical casing in a halfspace. An opportunity for future research is to explore strategies for addressing the ``upscaling'' problem and capturing the impacts due to permeability on a coarser scale for 3D simulations. Another complicating factor is that often magnetic permeability is unknown, so another avenue of future research is to develop strategies to develop an approach for estimating permeability from EM data.

The ability to perform numerical simulations and collect high-quality data continues to improve, and this opens up opportunities to increase the utility of electromagnetics in applications where signals may be subtle or the settings complex. Understanding the details of what contributes to an EM response will be important for extracting insights from those data. We hope that our work contributes to that understanding and helps in the utilization of EM methods in settings with steel infrastructure.

\section{Data availability}

The code used is available at \texttt{https://doi.org/10.5281/zenodo.7723282}.

\section{Acknowledgements}
The authors are grateful to Seogi Kang. Much of our analysis was inspired by the course material prepared for the Society of Exploration Geophysics Distinguished Instructor Short Course (DISC) in 2017 for which Seogi was a co-instructor. The authors are also grateful to the SimPEG community for contributions and continued maintenance of the code-base.

We are grateful to Dr. Michael Wilt for his review and constructive commentary which greatly improved our manuscript.

\clearpage

\bibliographystyle{gji}
\bibliography{bibliography}

\clearpage

\appendix

\section{Analysis of the frequency domain EM response of a conductive, permeable well}
\label{app:fdem}

\subsection{Currents in the formation}
\begin{figure}
    \begin{center}
    \includegraphics[width=\textwidth]{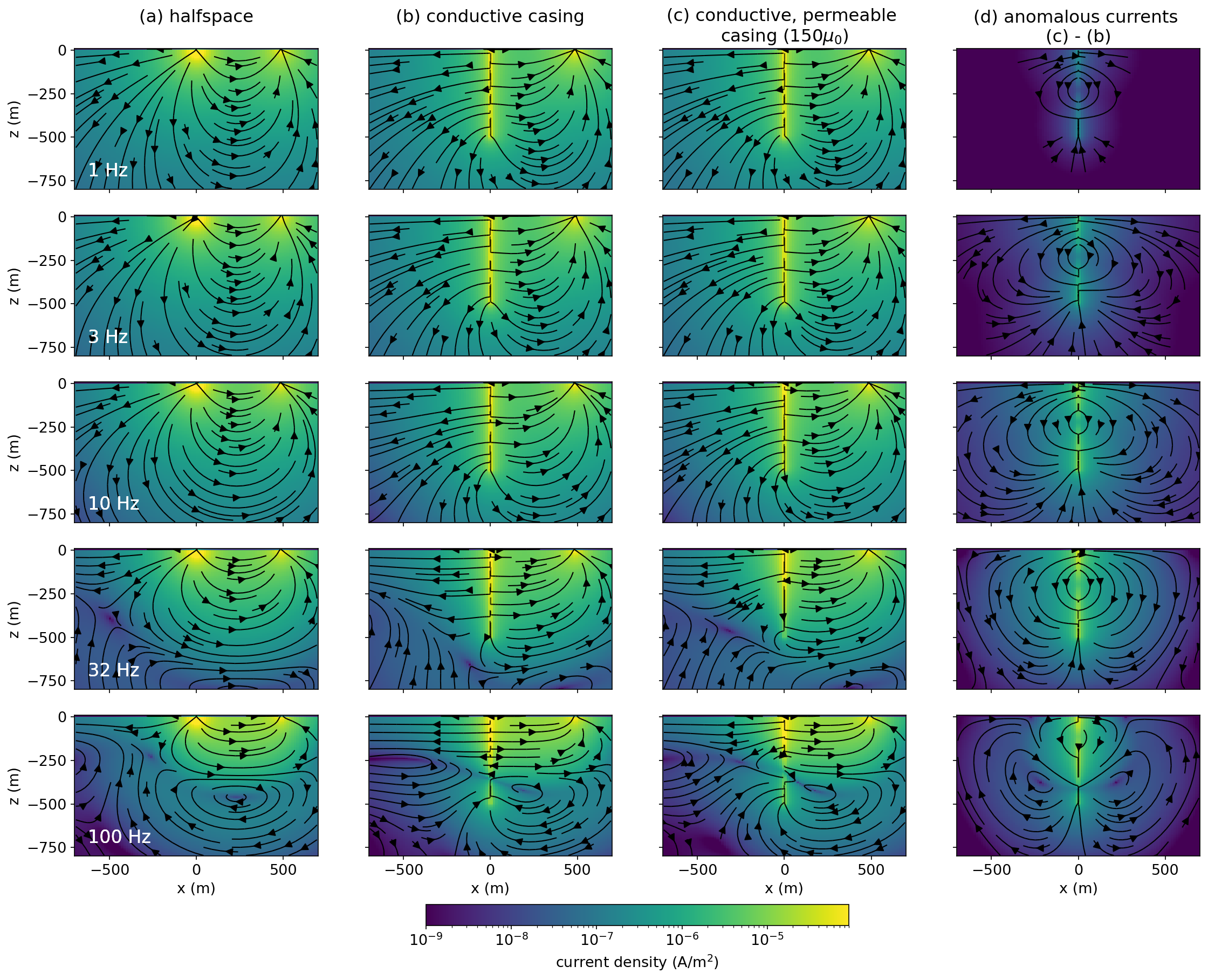}
    \end{center}
\caption{Cross sections of the real component of the current density in frequency domain EM experiments over (a) a half-space, (b) a conductive well ($5\times10^6$ S/m, $\mu_0$), and (c) a conductive, permeable well ($5\times10^6$ S/m, $150\mu_0$). The panel on the right shows (d) the difference between the conductive, permeable well and the conductive well.
}
\label{fig:fdem-cross-section-currents-real}
\end{figure}

We consider the EM response of a conductive, permeable casing in a frequency domain EM experiment. We use the same survey geometry and physical properties as shown in Figure \ref{fig:setup}, and now consider a harmonic waveform. We again begin by examining the currents in the subsurface and show the real component of the currents in Figure \ref{fig:fdem-cross-section-currents-real} for (a) a halfspace model, (b) a conductive casing ($\sigma = 5\times10^6$ S/m, $\mu=\mu_0$), (c) a conductive, permeable casing ($\sigma = 5\times10^6$ S/m, $\mu=150\mu_0$), and (d) the anomalous currents due to the permeability of the casing. For frequencies up to 10Hz, the response is visually identical to the DC response we observed in Figure \ref{fig:tdem-cross-section-currents}. This is to be expected, at the resistive limit of Maxwell's equations, the EM response is dominated by the galvanic currents. At higher frequencies, we can see the effects of induction for all of the models with the development of current systems at depth that circulate in a direction opposite to the galvanic currents.

\begin{figure}
    \begin{center}
    \includegraphics[width=\textwidth]{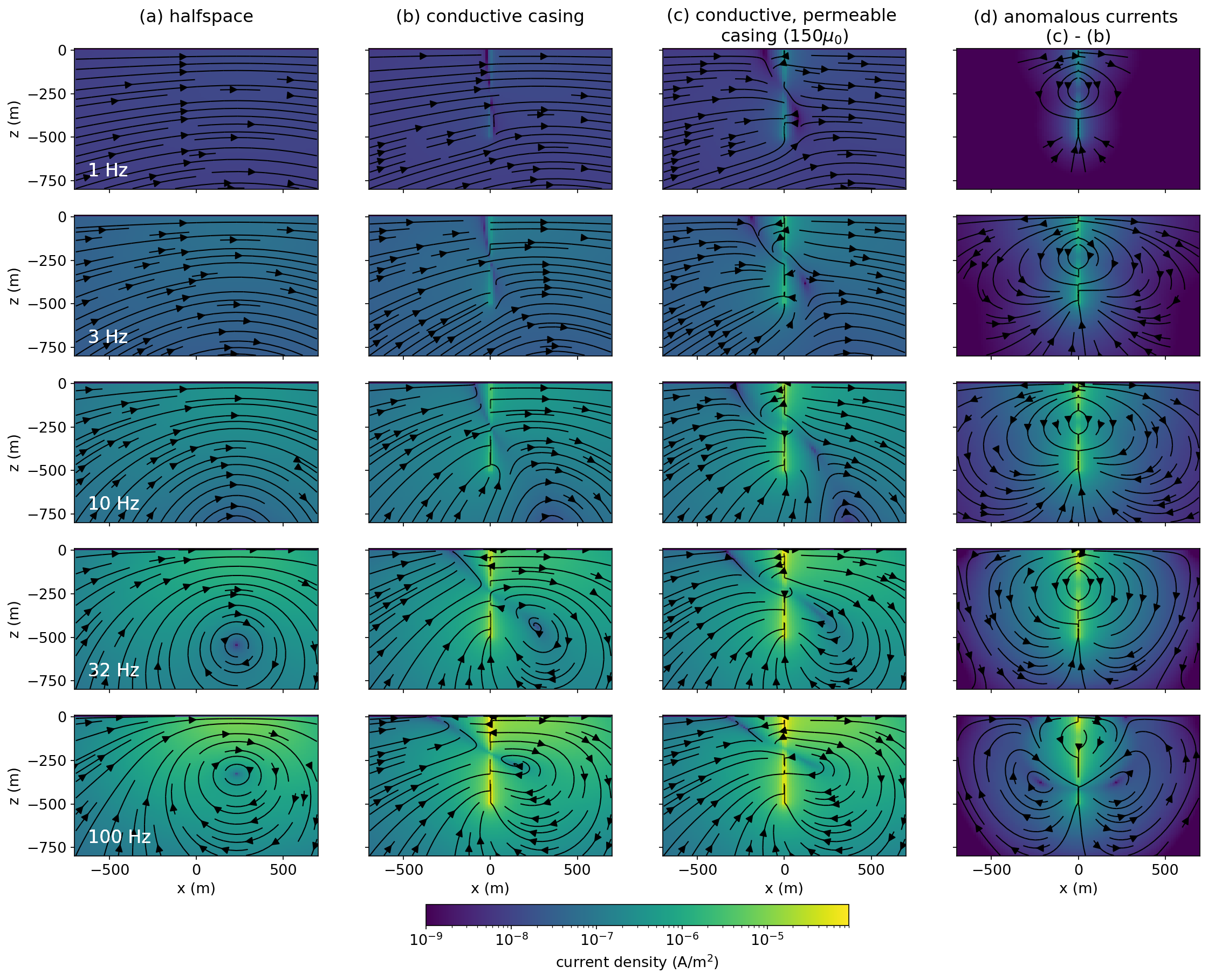}
    \end{center}
\caption{
    Cross sections of the inductive part of the real current density (real component of the current density minus the DC current density).
}
\label{fig:fdem-cross-section-currents-subtract-dc}
\end{figure}

A benefit of working in the frequency domain as compared to the time domain is that it is simpler to isolate the inductive component of a response. In the time domain, after the transmitter is turned off, both the galvanic and image current-systems diffuse downwards and outwards through time, so there is no straightforward way to separate them. However, in the frequency domain, where the transmitter is always on, the geometry of the galvanic portion of the response remains the same at all frequencies. Thus, we can isolate the inductive response in the real component by subtracting off the response at the DC limit (0 Hz). This is what is shown in Figure \ref{fig:fdem-cross-section-currents-subtract-dc}.

Isolating the inductive component of the real response provides a useful connection to our analysis in the time-domain. The geometry of currents resembles those observed in the time-domain, where time and frequency are inversely related. Note that the arrows are in the opposite direction; this is a function of the choice of an $e^{i\omega t}$ Fourier transform convention that follows \citep{Ward1988}. Our choice of Fourier transform convention is somewhat inconvenient for making comparisons between the time and frequency domains, because of the sign reversal, however, we chose to use this convention because it is common in geophysics and is the convention that is implemented in SimPEG.

\begin{figure}
    \begin{center}
    \includegraphics[width=\textwidth]{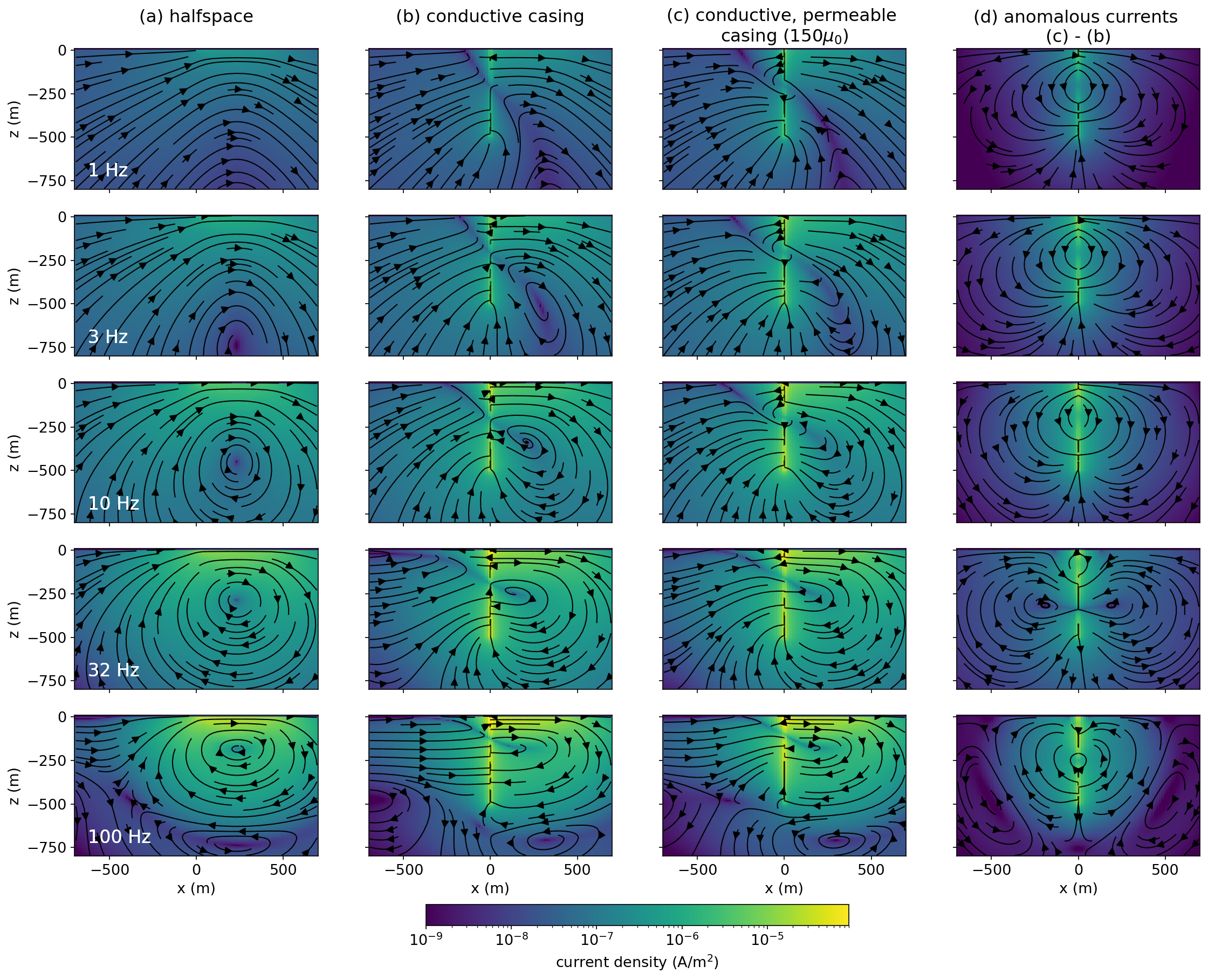}
    \end{center}
\caption{
    Cross sections of the imaginary component of the current density.
}
\label{fig:fdem-cross-section-currents-imag}
\end{figure}

In Figure \ref{fig:fdem-cross-section-currents-imag}, we show the imaginary component of the currents. The geometry is similar to the inductive component of the real response in Figure \ref{fig:fdem-cross-section-currents-subtract-dc}, but in the imaginary component, we also see the influence of the magnetic field. The ``flattening'' of the currents beneath the transmitter wire (x=0m to x=1000m) is a result of the rotational magnetic field created by the transmitter according to Ampere's law. In both Figures \ref{fig:fdem-cross-section-currents-subtract-dc} and \ref{fig:fdem-cross-section-currents-imag}, we see the ``shadow-zone'' that is responsible for the zero-crossing in the imaginary component that we now understand to be the interaction of the image currents with the channelled currents due to the casing.

\subsection{Impacts of permeability on excitation}
\begin{figure}
    \begin{center}
    \includegraphics[width=\textwidth]{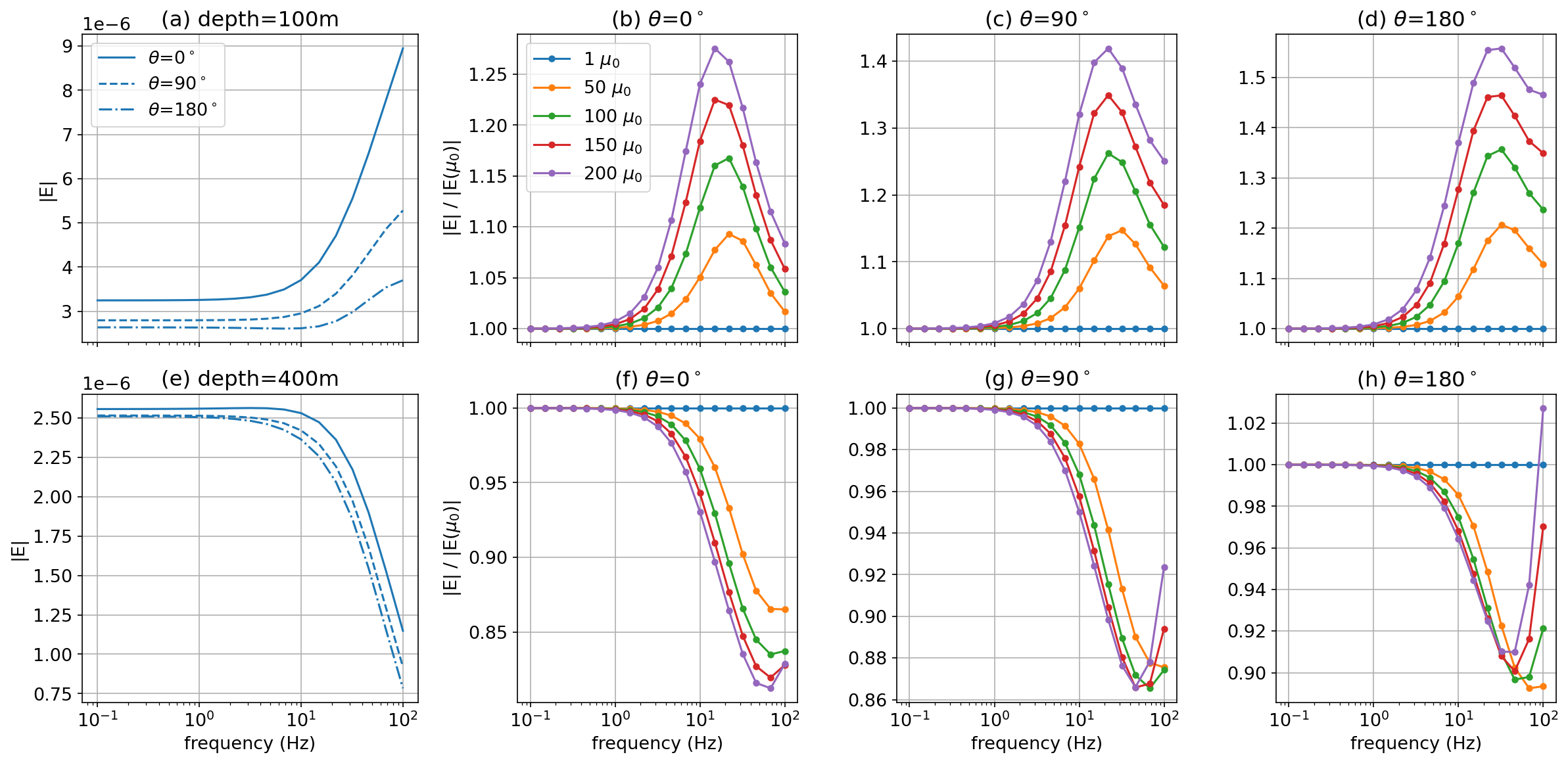}
    \end{center}
\caption{
    Amplitude of the average electric field over a test volume that is the same as was used in Figure \ref{fig:excitation-time-integrated}. The different line-styles in (a) and (e) indicate different azimuths for the non-permeable well ($\mu_0$). Panels (b), (c), and (d) show the ratio of the amplitude of the electric field with respect to the non-permeable well ($\mu=\mu_0$) for the test volume centered at a depth of 100m. Panels (f), (g) and (h) show the ratios of the electric field amplitude for the test volume centered at 400m depth.
}
\label{fig:excitation-freq}
\end{figure}

To examine how permeability impacts our ability to excite a target, we repeat the same analysis as was done for Figure \ref{fig:excitation-time-integrated}. We compute the amplitude of the average electric field in a test volume that is offset from the well. This is shown in Figure \ref{fig:excitation-freq}. In the top row, we consider a test volume centred at 100m depth, and in the bottom row, we consider a test volume centered at 400m depth. In (a) and (d), we show the amplitude of the electric field for a non-permeable casing at three different azimuths relative to the transmitter wire (indicated by the different line-styles). In (b-d) and (f-h) we show the ratio of the amplitude of the electric field with respect to that due to the non-permeable well. As we might expect, at frequencies below 1 Hz, there is minimal influence of permeability, but as the frequency increases, we begin to see the impacts of permeability. For a target near the surface, permeability enhances the electric field in the target volume. Focusing on the well with a permeability of 150$\mu_0$, we see that near 20 Hz, the amplitude of the electric field is increased by a factor of 20\% to $>$40\% depending on the azimuth. Even for 5 Hz, the enhancement of the electric field is substantial, ranging from 7\% and 9\%.

\begin{figure}
    \begin{center}
    \includegraphics[width=\textwidth]{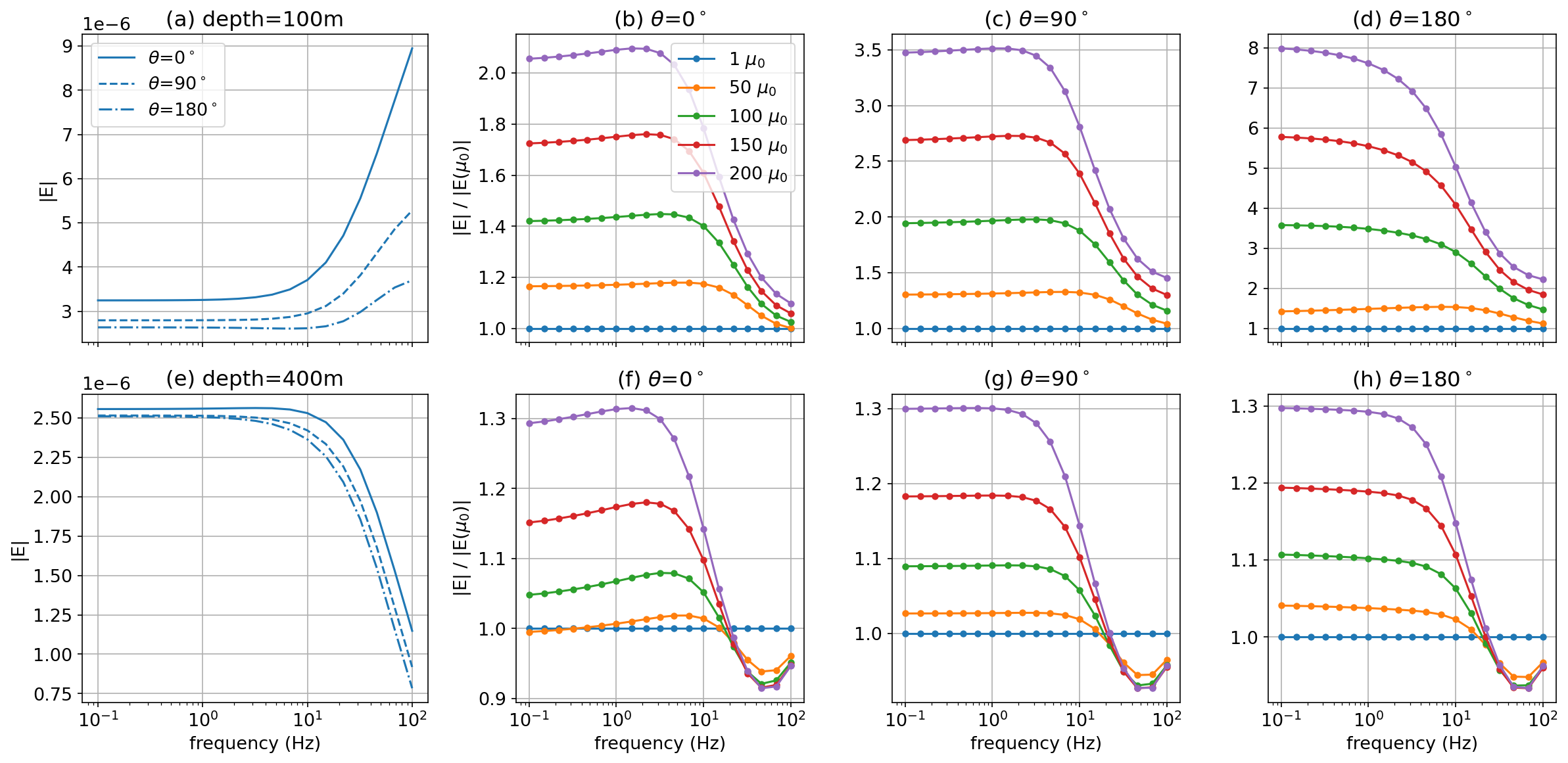}
    \end{center}
\caption{
    Amplitude of only the inductive part of the electric field (total minus DC) over a test volume for the same experiment as shown in Figure \ref{fig:excitation-freq}.
}
\label{fig:excitation-freq-subtract-dc}
\end{figure}

Interestingly, for a deeper test volume, permeability has the opposite effect and reduces the average electric field. This is consistent with the currents shown in Figure \ref{fig:fdem-cross-section-currents-real} where we see the total currents point away from the well at depth, while the anomalous currents in \ref{fig:fdem-cross-section-currents-real}(d) point towards the well. At the highest frequencies, there are reversals of currents at depth that can be seen at 100 Hz in Figure \ref{fig:fdem-cross-section-currents-real}. When the direction of the total and anomalous currents are the same, the response can then be enhanced, as seen at 100Hz in Figure \ref{fig:excitation-freq}(h). A reduction in the amplitude of the electric field as the permeability of the well is increased was not observed in the time-domain (Figure \ref{fig:excitation-time-integrated}). If, however, we subtract the galvanic response from the frequency domain results and isolate the inductive response, we obtain the results in Figure \ref{fig:excitation-freq-subtract-dc}. At both depths, there is an enhancement of the average electric field for frequencies below 20 Hz. This is consistent with the currents and anomalous currents in Figures \ref{fig:fdem-cross-section-currents-subtract-dc} and \ref{fig:fdem-cross-section-currents-imag}. Permeability can enhance the inductive component of the EM response, but whether this translates to an overall increase or decrease in the amplitude fields depends upon both the galvanic and induced components.

\subsection{Currents within the casing}
\begin{figure}
    \begin{center}
    \includegraphics[width=1\textwidth]{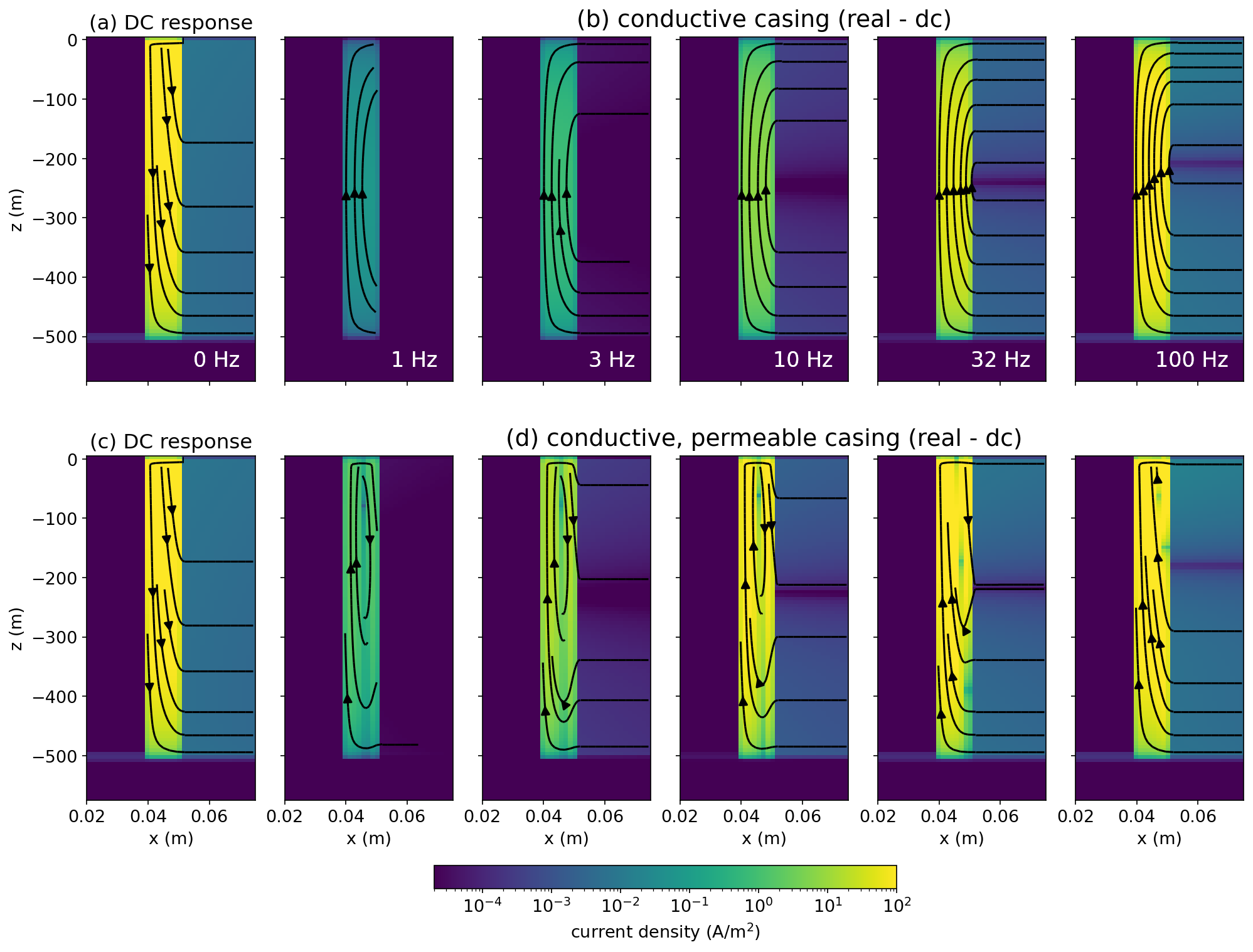}
    \end{center}
\caption{
    Cross sections of DC current density within (a) a conductive casing ($5\times10^6$ S/m, $\mu_0$) and (c) a conductive, permeable casing ($5\times10^6$ S/m, $150\mu_0$). The cross sections in (b) and (d) show the inductive part of the real current density (total - DC).
}
\label{fig:fdem-casing-currents-real}
\end{figure}

\begin{figure}
    \begin{center}
    \includegraphics[width=0.8\textwidth]{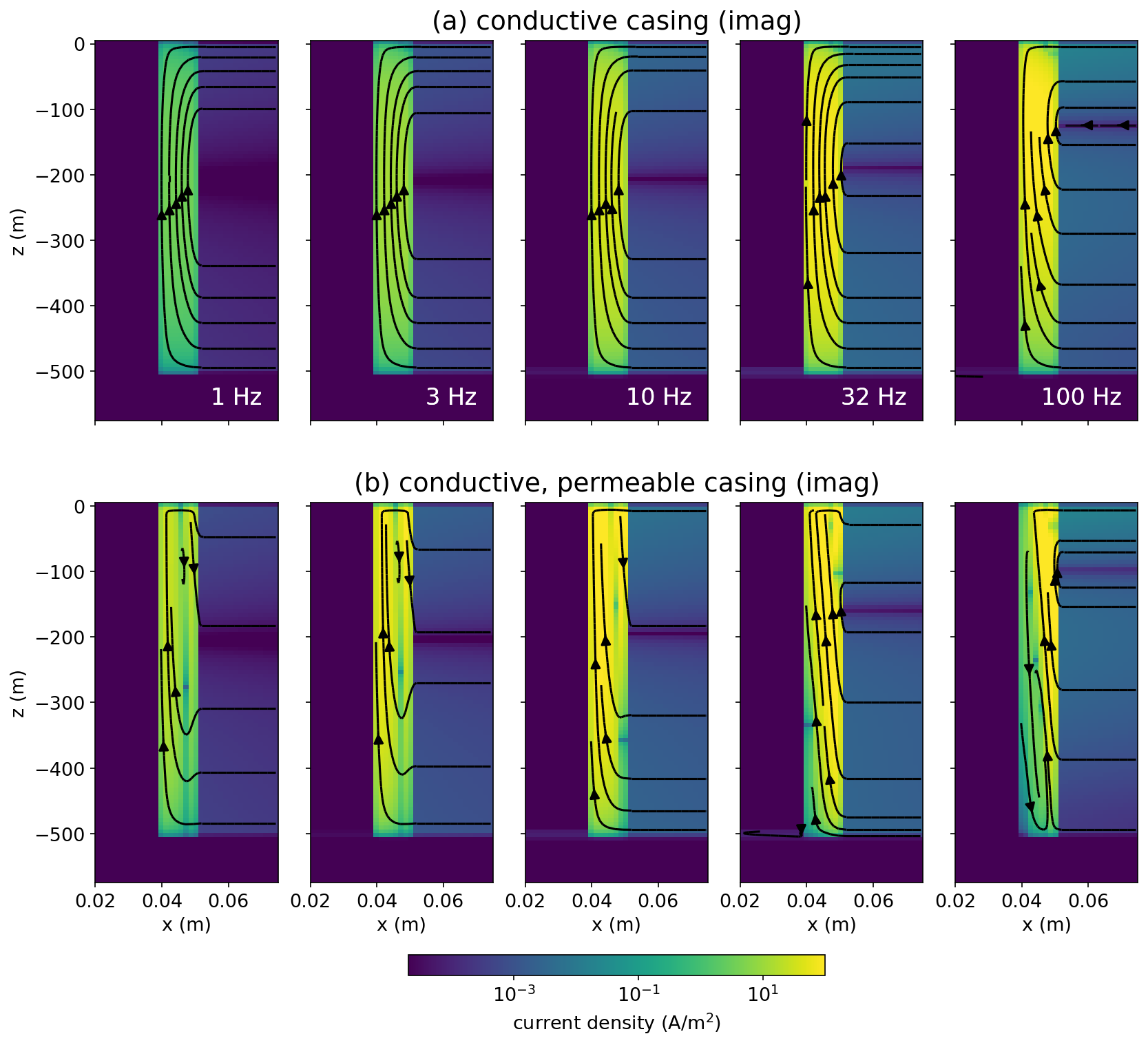}
    \end{center}
\caption{
    Cross sections of the imaginary part of the current density within (a) a conductive casing ($5\times10^6$ S/m, $\mu_0$) and (b) a conductive, permeable casing ($5\times10^6$ S/m, $150\mu_0$).
}
\label{fig:fdem-casing-currents-imag}
\end{figure}

To visualize the impacts of permeability on the currents in the casing, we zoom in to the casing wall and plot the currents at a range of frequencies. In Figure \ref{fig:fdem-casing-currents-real} we show (a) the DC response for a conductive casing, and (b) the inductive component of the real response, computed by subtracting the DC currents from the real component of the total currents at each frequency. The bottom row similarly shows the real component of the currents for a conductive, permeable casing ($\mu_r = 150$). Figure \ref{fig:fdem-casing-currents-imag} shows the imaginary component of the currents.

In comparing with the time-domain images in Figure \ref{fig:tdem-casing-currents}, we see that high-frequencies in the inductive part of the real component (Figure \ref{fig:fdem-casing-currents-real}) are similar in geometry and amplitude to the early times. The response at low frequencies is similarly analogous to later times, and we can again observe the poloidal current system in the permeable well. Again note that directions are opposite to the time domain because of the Fourier transform convention.

\subsection{Explaining the poloidal current system}
The same reasoning as we used to explain the poloidal current system in the time domain holds for the frequency domain. Replacing the time-domain fields and fluxes in equation \ref{eq:permeability-ampere-no-source} with frequency domain variables, we have
\begin{equation}
\nabla \times \vec{B} = \nabla \ln \mu_r \times \vec{B} + \mu\sigma\vec{E}
\label{eq:permeability-ampere-no-source-freq}
\end{equation}

\begin{figure}
    \begin{center}
    \includegraphics[width=\textwidth]{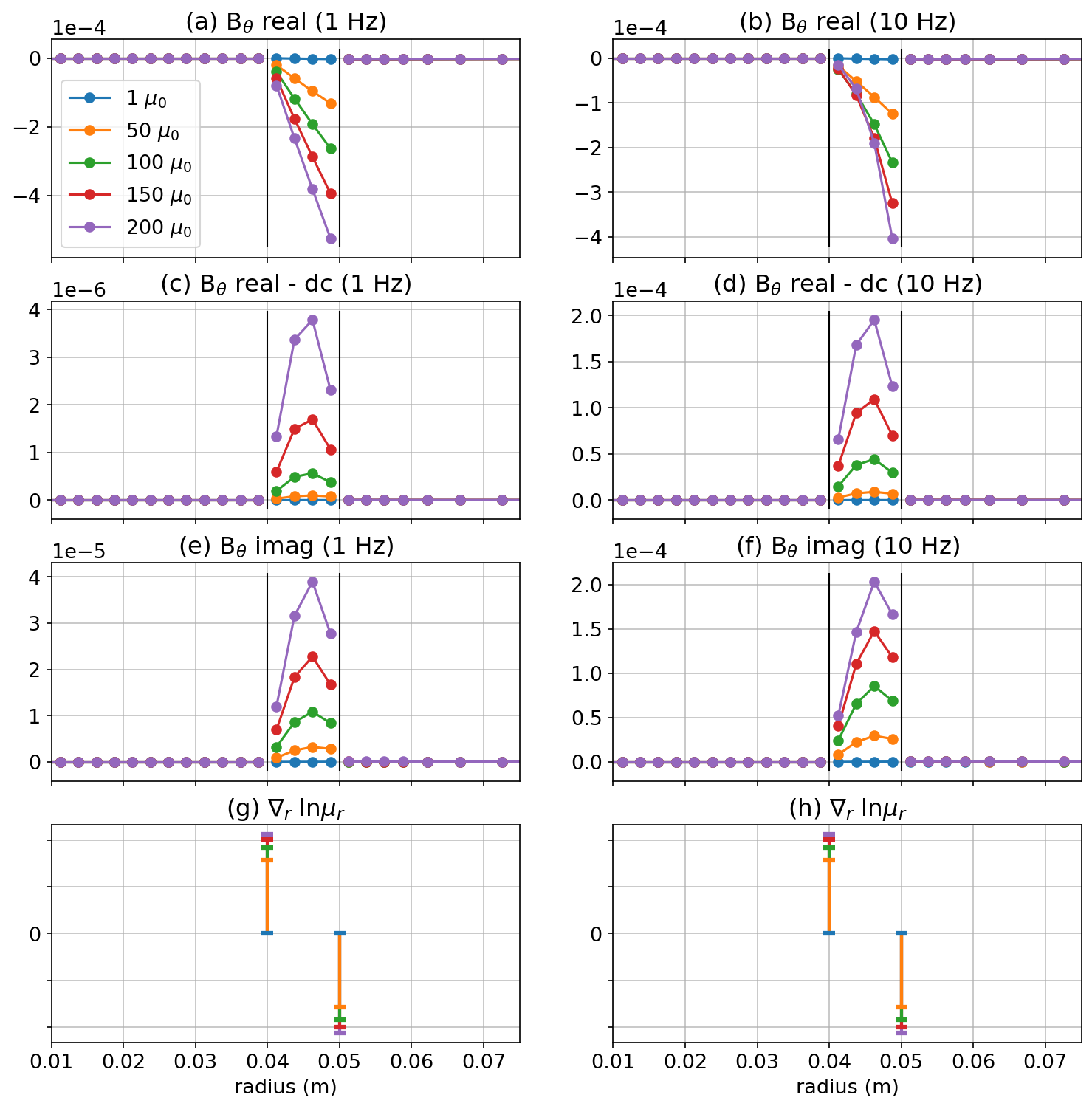}
    \end{center}
\caption{
    (a-b) Real part of the azimuthal component of the magnetic flux at 1 Hz and 10 Hz, respectively.
    (c-d) Inductive part of the real component of the magnetic flux (real - dc).
    (e-f) Imaginary component of the magnetic flux.
    (g-h) Radial component of $\nabla \ln \mu_r$.
}
\label{fig:magnetization-term-freq}
\end{figure}

In Figure \ref{fig:magnetization-term-freq}, we plot the real and imaginary component of the magnetic flux density in the vicinity of the casing at two different frequencies 1 Hz and 10 Hz. Examining equation \ref{eq:permeability-ampere-no-source-freq}, we know that at DC (0 Hz), there is no influence of $\vec{B}$ on $\vec{E}$. So although the magnetization term in equation \ref{eq:permeability-ampere-no-source-freq} is non-zero when the casing is permeable, this portion of the response has no influence on the currents or electric fields (by Faraday's law). So we subtract off the DC response and focus our attention on the inductive parts of $\vec{B}_\theta$ in Figure \ref{fig:magnetization-term-freq} (c-f) to understand how the permeability of the casing impacts the EM response. The real, inductive part of the magnetic flux density (c-d) is positive along the outer edge of the casing and therefore contributes to a downward-going magnetization current. This is similarly true for the imaginary component.

\end{document}